\documentclass[a4paper,11pt]{article}

\usepackage[utf8]{inputenc}
\usepackage[T1]{fontenc}
\usepackage[english]{babel}

\usepackage{indentfirst}
\usepackage{amsmath}
\usepackage{amsthm}
\usepackage{amssymb}
\usepackage{graphicx}
\usepackage[font=small]{caption}
\usepackage{subcaption}
\usepackage{psfrag}
\usepackage{color}
\usepackage{pgf,pgfnodes}

\allowhyphens

\newcommand{\complex}{\mathbb{C}}

\newcommand{\eps}{\varepsilon}

\newcommand{\ordo}{\mathcal{O}}

\newcommand{\ket}[1]{{\left|#1\right\rangle}}
\newcommand{\bra}[1]{{\left\langle #1\right|}}

\newcommand{\skalarszorzat}[2]{{\langle #1 | #2 \rangle}}

\newcommand{\vev}[1]{\left\langle #1 \right\rangle}

\usepackage{ifpdf}

\ifpdf
\usepackage{epstopdf}
\usepackage[pdftex,colorlinks,urlcolor=blue,citecolor=blue,linkcolor=blue]{hyperref}
\else
\usepackage[hypertex,colorlinks,urlcolor=blue,citecolor=blue,linkcolor=blue]{hyperref}
\fi

\makeatother

\begin{document}

\numberwithin{equation}{section}

\title{Quantum quenches and Generalized Gibbs Ensemble in a \\ Bethe
  Ansatz solvable lattice model of interacting
  bosons}

\author{Bal\'azs Pozsgay$^1$\\
~\\
 $^{1}$MTA--BME \textquotedbl{}Momentum\textquotedbl{} Statistical
Field Theory Research Group\\
1111 Budapest, Budafoki út 8, Hungary
}

\maketitle
\abstract{We consider quantum quenches in the so-called $q$-boson
  lattice model. We argue that the Generalized Eigenstate
  Thermalization Hypothesis holds in this model, therefore the Generalized Gibbs
  Ensemble (GGE) gives a valid description of the stationary states in
the long time limit. For a special class of initial states (which are
the pure Fock states in the local basis) we are able
to provide the GGE predictions for the resulting root densities. We
also give predictions for the long-time limit of certain local
operators. In the $q\to\infty$ limit the calculations simplify considerably, 
the wave functions are given by Schur polynomials and the overlaps
with the initial states can be written as 
simple determinants. In two cases we prove rigorously that the GGE
prediction for the root density is correct. Moreover, we calculate the
exact time dependence of a physical observable (the one-site Emptiness
Formation Probability) for the quench starting
from the state with exactly one particle per site. In the long-time
limit the GGE prediction is recovered.
}

\section{Introduction}

The problems of equilibration and thermalization of closed quantum systems have
attracted considerable interest recently
\cite{Silva-quench-colloquium,huse-review}. One of 
the central questions is whether the principles of statistical physics can
be derived from the unitary time evolution of the quantum
system. Research in this field has been motivated partly by new 
experimental techniques (for example with cold atoms
\cite{cold-atom-review}) where 
an almost perfect isolation from the environment can be achieved, and
therefore equilibration induced by the system itself can be studied.

Equilibration in a quantum mechanical system means that the
expectation values of physical observables approach stationary values 
in the long time limit. 
Thermalization happens when these coincide with
predictions obtained from a thermal ensemble. 
One dimensional integrable models comprise a special class of systems which
possess a family of higher conserved charges in addition to the usual
ones. These extra conservation laws prevent thermalization in the
usual sense. Instead, it was proposed in \cite{rigol-gge} that the
stationary values of local observables should be described by the
Generalized Gibbs Ensemble (GGE). This ensemble includes all the charges
with Lagrange-multipliers fixed by the mean values of the charges in 
the initial state. 

Since its inception the idea of the GGE has attracted considerable
interest and sparked many discussions. 
A large body of numerical evidence for its validity was found in
the lattice model of hard-core bosons
\cite{rigol-gge,rigol-2,rigol-GETH,rigol-3} (see also \cite{rigol-anyons}) and
it was proven to be true in
free theories or models equivalent to free fermions
\cite{free-gge-1,free-gge-2,free-gge-3,free-gge-4,ising-quench-1,ising-quench-2,essler-truncated-gge,spyros-calabrese-gge}. 
However, it was found in \cite{JS-oTBA} that in the interacting
spin-1/2 XXZ chain the GGE (built on the strictly local charges) gives different predictions 
than the Quench Action (QA) method \cite{quench-action}, which (as opposed to
the GGE) is built on first principles and does not involve any
assumptions or approximations. 
Furthermore conclusive evidence was found
in a case of a specific quench problem in
\cite{sajat-oTBA} that while the predictions of the QA method coincide with
results of real-time simulations, the GGE predictions \cite{sajat-xxz-gge,essler-xxz-gge,fagotti-collura-essler-calabrese} are not correct.

It was argued in \cite{rigol-GETH} that equilibration to the GGE
predictions can be explained by the Generalized Eigenstate
Thermalization Hypothesis (GETH), which roughly states that if there are two
states which have almost the same mean values of the conserved charges, then local
correlations in the two states should be also close to each other. If the
GETH holds than the GGE is valid for quenches from any initial state satisfying the
cluster decomposition principle. It was recently shown in
\cite{sajat-GETH} that the failure of the GGE in the XXZ spin chain
can be attributed to the failure of the GETH. It was argued in
\cite{sajat-GETH,andrei-gge} that this is a generic property of
integrable models with multiple particle species.

The question remains whether the GETH and the GGE can be correct in any
genuinely interacting integrable model. To find an example one should
certainly look for models with one particle type. An obvious choice
would be the Lieb-Liniger (LL) model, which is a continuum theory of
1D interacting bosons \cite{Lieb-Liniger}. Quenches in the LL model
have been investigated in a number of papers recently 
\cite{jorn-ll,caux-konik-ll,andrei-ll-1,marci-ll-quench1,marci-ll-quench2,caux-stb-LL-BEC-quench,andrei-ll-2,andrei-ll-3,andrei-ll-4}. However,
it was found in \cite{marci-ll-quench1} that for an interaction quench
from zero to finite coupling the expectation values of the higher
charges are divergent, and therefore the GGE can not be defined in
that case. The problem was circumvented 
by applying a lattice regularization using the so-called
$q$-boson model \cite{q-bozon-eredeti,q-bozon-eredeti-a,q-bozon-eredeti2}. The QA solution of
this interaction quench was later given in
\cite{caux-stb-LL-BEC-quench}, where it was argued that the GGE can
not be correct in the LL model due to the aforementioned divergences and the
observed logarithmic singularities in the Bethe root densities, which
can not be captured by the GGE.

In this paper we investigate quantum quenches in the $q$-boson model,
without the goal of taking the continuum limit towards the Bose
gas. In this lattice model 
there is only one particle type in the spectrum, and the
infinities encountered in the LL model do not appear here. 
Therefore it is an ideal testing ground for the GETH and the GGE.

The paper is organized as follows. In Section \ref{sec:BA} we review
the Bethe Ansatz solution of the model and the construction of the
higher conserved charges. In \ref{sec:GGE} we construct the GGE
density matrix for this model and argue that the GETH holds, therefore
the GGE is valid. In \ref{sec:120} we consider specific quench
problems and provide the GGE predictions for a class of initial states.
The $q\to\infty$ limit of the model is investigated in
\ref{sec:infty1}, where the equilibrium properties of the model are established.
Quenches in the $q\to\infty$ limit are investigated in
\ref{sec:infty}, 
where we confirm the GGE predictions in two simple
cases by analyzing the exact overlaps. Also, we derive an analytic
formula for the time dependence of a simple physical observable, the
one-site Emptiness Formation Probability. The long-time limit of this
quantity is found to agree with the GGE prediction. The $q=1$ limit of
the model (the case of free bosons) is considered in
\ref{sec:bosons}. Finally, Section \ref{sec:vege} includes our
conclusions, a number of remarks about our results, and a list of open problems.

\section{The model and its Bethe Ansatz solution}

\label{sec:BA}

Consider a lattice consisting of $L$ sites such that the
configuration space of each site is a single bosonic space. Let
us define the canonical Bose operators $b_j$, $b^\dagger_j$,
$N_j$ acting on site $j$ by the usual commutation relations
\begin{equation*}
  [b_j,b_k^\dagger]=\delta_{j,k} N_k\qquad 
[N_j,b_k]=-\delta_{j,k}  b_k\qquad
[N_j,b_k^\dagger]=-\delta_{j,k}  b_k^\dagger.
\end{equation*}
The action of these operators on the local states $\ket{n}_j$,
$n=0\dots\infty$ is given by
\begin{equation*}
  b_j\ket{n}_j=\sqrt{n}\ket{n-1}_j\qquad
 b_j^\dagger\ket{n}_j=\sqrt{n+1}\ket{n+1}_j\qquad
 N_j\ket{n}_j=n\ket{n-1}_j.
\end{equation*}
We also define the operators $B_j^\dagger$, $B_j$  by their action
\begin{equation*}
  B_j\ket{n}_j=\sqrt{[n]_q}\ket{n-1}_j\qquad
 B_j^\dagger\ket{n}_j=\sqrt{[n+1]_q}\ket{n+1}_j,
\end{equation*}
where
\begin{equation*}
  [x]_q=\frac{1-q^{-2x}}{1-q^{-2}}.
\end{equation*}
The parameter $q$ is an arbitrary real number. In the present work we
will consider the cases $q\ge 1$ and we will use the parametrization
$q=e^{\eta}$, $\eta>0$. It is easy to check that the following
commutation relations hold:
\begin{equation*}
[N_k,B_k]=-  B_k\qquad
[N_k,B_k^\dagger]=-  B_k^\dagger\qquad
[B_k,B_k^\dagger]= q^{-2N_k}
\end{equation*}
These equations are the defining relations of the so-called $q$-boson
algebra \cite{q-bozon-algebra}. The canonical Bose
operators are recovered in the $q\to 1$ limit:
\begin{equation*}
  \lim_{q\to 1} B_k=b_k\qquad
  \lim_{q\to 1} B_k^\dagger=b_k^\dagger.
\end{equation*}

The $q$-boson Hamiltonian is defined as
\begin{equation}
  \label{Hq}
H=-\sum_{j=1}^L (B_j^\dagger B_{j+1}+ B_{j+1}^\dagger B_{j}-2N_j),
\end{equation}
where periodic boundary conditions are assumed. 
Even though the Hamiltonian \eqref{Hq} has the form of a free
hopping model, there are interactions between the particles due to the
fact that the $B$ and $B^\dagger$ are not the canonical Bose
operators and the hopping amplitudes depend on the
local occupation numbers. The model can serve as a lattice
regularization of the 
Lieb-Liniger model
\cite{q-bozon-eredeti,q-bozon-eredeti-a,q-bozon-eredeti2,zvonarev-g3,marci-ll-quench1};
however, in the 
present work we focus on the lattice model only.

The $q$-boson Hamiltonian was solved in \cite{q-bozon-eredeti}
 by the Algebraic Bethe Ansatz (ABA). The coordinate Bethe
Ansatz wave functions were later calculated in \cite{q-bozon-coo-BA};
they are given by Hall-Littlewood functions. Here we review the (ABA)
solution, our exposition follows that of  \cite{zvonarev-g3}. 

\bigskip

Let us consider an auxiliary space
$V=\complex^2$ and define the so-called Lax operator, which is a matrix in
auxiliary space with matrix elements being operators in the bosonic
Fock spaces:
\begin{equation*}
L(\lambda)=
\begin{pmatrix}
 e^\lambda & \chi B^\dagger \\
\chi B & e^{-\lambda} 
\end{pmatrix}.
\end{equation*}
Here $\lambda$ is the rapidity parameter and $\chi^2=1-q^{-2}$. The
Lax operator satisfies the Yang-Baxter equation
\begin{equation}
\label{RLL}
 R(\lambda-\mu) \big(L(\lambda)\otimes L(\mu)\big)=
\big(L(\mu)\otimes L(\lambda)\big)  R(\lambda-\mu)
\end{equation}
with the $R$-matrix
\begin{equation*}
  R(u)=
  \begin{pmatrix}
    \sinh(u+\eta) & 0 & 0 & 0 \\
0 & \sinh(\eta) & q\sinh(u) & 0 \\
0 & q^{-1}\sinh(u) & \sinh(\eta) & 0 \\
0 & 0 & 0 & \sinh(u+\eta) \\
  \end{pmatrix}.
\end{equation*}
The central object of the ABA is the monodromy matrix, which is
given by
\begin{equation}
  \label{eq:T}
 T(\lambda)
=L_L(\lambda)L_{L-1}(\lambda)\dots L_1(\lambda) =
 \begin{pmatrix}
   A(\lambda) & B(\lambda) \\
C(\lambda) & D(\lambda) 
 \end{pmatrix}.
\end{equation}
It follows from \eqref{RLL} that the monodromy matrix satisfies the RTT-relation
\begin{equation}
\label{RTT}
 R(\lambda-\mu) \big(T(\lambda)\otimes T(\mu)\big)=
\big(T(\mu)\otimes T(\lambda)\big)  R(\lambda-\mu).
\end{equation}
A direct consequence of \eqref{RTT} is that the transfer matrix
defined as
\begin{equation*}
  \tau(\lambda)=\text{Tr } T(\lambda)=A(\lambda)+D(\lambda)
\end{equation*}
satisfies
\begin{equation}
\label{comm}
  [\tau(\lambda),\tau(\mu)]=0.
\end{equation}
It is easy to see that
\begin{equation}
\label{vegtelen}
  \lim_{\lambda\to -\infty} \Big(e^{L\lambda}\tau(\lambda)\Big)=1.
\end{equation}
The properties \eqref{comm} and \eqref{vegtelen} enable us to obtain a
commuting set of local charges by expanding the logarithm of the
transfer matrix around the point $\lambda= -\infty$. We define
\begin{equation}
\label{Idef}
  I_m=\frac{1}{(2m)!}\left(\frac{\partial}{\partial \xi}\right)^{2m}
\left.\log\left(\xi^L \tau(\xi)\right)\right|_{\xi=0},
\end{equation}
where $\xi=e^\lambda$. It follows from the definition of the transfer
matrix and the form of the Lax operator that $I_m$ is a sum of local
operators which span at most $m+1$ sites. The first two examples are
\begin{equation}
\label{charges}
\begin{split}
I_1&=\chi^2 \sum_j B_j^\dagger B_{j+1}\\  
I_2&=\chi^2(1-\frac{\chi^2}{2}) \sum_j(
 B_j^\dagger B_{j+2}-\frac{\chi^2}{2-\chi^2}B_j^\dagger B_j^\dagger
 B_{j+1}  B_{j+1}-\chi^2 B_j^\dagger B_{j+1}^\dagger
 B_{j+1}  B_{j+2}).
\end{split}
\end{equation}
A formula for $I_3$ is also given in \cite{zvonarev-g3}.

These operators are not Hermitian. It is useful to define the charges
with negative indices as their adjoint:
\begin{equation*}
  I_{-n}=(I_n)^\dagger.
\end{equation*}
They can be obtained by expanding the transfer matrix around $\lambda=\infty$.

The particle number operator
\begin{equation*}
  N=\sum_j N_j
\end{equation*}
commutes with all of the charges, which follows from the fact that the
transfer matrix only includes terms with an equal number of
$B^\dagger$ and $B$ operators. We define
$I_0\equiv N$. The Hamiltonian can then be written as
\begin{equation*}
  H=-\frac{I_1+I_{-1}}{\chi^2}+2I_0.
\end{equation*}

Eigenstates of the system are constructed using the $B$-operators of
the monodromy matrix:
\begin{equation}
\label{Bstate}
  \ket{\{\lambda\}_N}=\prod_{j=1}^N B(\lambda_j) \ket{0},
\end{equation}
where $\ket{0}$ is the Fock vacuum. The parameters $\lambda_j$ are the
rapidities of the interacting bosons. A state of the form
\eqref{Bstate} is an eigenstate of the transfer matrix if the
rapidities satisfy the Bethe equations:
\begin{equation}
  \label{Be}
  e^{2L\lambda_j}\prod_{k\ne j} 
\frac{\sinh(\lambda_j-\lambda_k+\eta)}{\sinh(\lambda_j-\lambda_k-\eta)}=1.
\end{equation}
The eigenvalues of the transfer matrix on the Bethe states are
\begin{equation}
\label{tau}
 \tau(u)  \ket{\{\lambda\}_N}=
\frac{1}{q^N}\left(
e^{Lu} \prod_{j=1}^N f(u,\lambda_j)+e^{-Lu} \prod_{j=1}^N f(\lambda_j,u)
\right) 
\ket{\{\lambda\}_N},
\end{equation}
where 
\begin{equation*}
  f(u)=\frac{\sinh(u+\eta)}{\sinh(u)}
\end{equation*}
Eigenvalues of the local charges are easily obtained using the
definition \eqref{Idef}. 
It is easy to see that they can be expressed as sums of single
particle eigenfunctions:
\begin{equation}
\label{hahaha}
  I_m\ket{\{\lambda\}_N} =
\sum_{j=1}^N i_m(\lambda_j),
\end{equation}
where
\begin{equation*}
  i_m(\lambda) =\frac{1}{(2m)!}\left(\frac{\partial}{\partial \xi}\right)^{2m}
\left.\log\left(f(\lambda,\log(\xi))\right)\right|_{\xi=0}.
\end{equation*}
In \eqref{hahaha} we used that the charge $I_m$ only exists in
lattices with $L>m$, therefore it is enough to keep the second term
from \eqref{tau}. Using the substitution $e^\lambda=a$ the derivatives are
calculated easily:
\begin{equation*}
\begin{split}
   i_m(\lambda) 
 &=\frac{1}{(2m)!}\left(\frac{\partial}{\partial \xi}\right)^{2m}
\left.\left[\log(1-\xi^2/(a^2q^2))-\log(1-\xi^2/a^2)
\right]
\right|_{\xi=0}\\
&=\frac{1}{m} \left(-\frac{1}{(aq)^{2m}}+\frac{1}{a^{2m}}\right)=
\frac{1}{m}(1-q^{-2m}) e^{-2m\lambda}.
\end{split}
\end{equation*}

It is useful to parametrize the rapidities as $\lambda=ip/2$. This way
the Bethe equations take the form
\begin{equation}
  \label{Be2}
  e^{ip_jL}\prod_{k\ne j} 
\frac{\sin((p_j-p_k)/2-i\eta)}{\sin((p_j-p_k)/2+i\eta)}=1.
\end{equation}
In the case of $\eta>0$ considered in the present work all solutions
to \eqref{Be2} are real numbers and they can be chosen to lie in the interval $[-\pi,\pi]$.

In terms of the $p$-variables the single particle eigenvalues of the charges take the form
\begin{equation*}
   i_m(p) =\frac{1}{|m|}(1-q^{-2|m|}) e^{-imp}.
\end{equation*}
The single particle energy is
\begin{equation*}
  e(p)=4\sin^2(p/2).
\end{equation*}
The $p$-variables are the physical pseudo-momenta on the lattice,
because the single particle eigenvalue for the translation by one-site is $e^{ip}$.

For the sake of completeness we note that the local $q$-boson
operators at sites $1$ and $L$ can be reconstructed from the
off-diagonal elements of the monodromy matrix as
\begin{equation}
\label{inverseproblem1}
  \begin{split}
    \lim_{\lambda\to\infty} \left(e^{-(L-1)\lambda} B(\lambda)\right)
= \chi B_1^\dagger\qquad&\qquad
    \lim_{\lambda\to-\infty} \left(e^{(L-1)\lambda} B(\lambda)\right)
= \chi B_M^\dagger\\
    \lim_{\lambda\to \infty} \left(e^{-(L-1)\lambda} C(\lambda)\right)
= \chi B_M\qquad&\qquad
    \lim_{\lambda\to -\infty} \left(e^{(L-1)\lambda} C(\lambda)\right)
= \chi B_1.
  \end{split}
\end{equation}
However, there are no such formulas for the other local $q$-boson
operators and the general solution of the so-called ``quantum inverse problem''
\cite{goehmann-korepin-inverse,maillet-terras-inverse} is not known. 

\subsection{Thermodynamic limit}

We will be interested in physical situations where there is a large
number of particles in a large volume such that the particle density
is finite. As usually we introduce the densities of Bethe roots
$\rho_r(p)$ and holes $\rho_h(p)$ such that in a large 
volume the total particle density is given by
\begin{equation*}
  \frac{N}{L}=\int_{-\pi}^\pi \frac{dp}{2\pi}\rho_r(p).
\end{equation*}
It follows from the Bethe equations that 
\begin{equation}
\label{rhoegy}
  \rho_r(p)+\rho_h(p)=1+\int \frac{du}{2\pi} \varphi(p-u) \rho_r(u)
\end{equation}
with
\begin{equation}
\label{vp}
\begin{split}
\varphi(u)&=\frac{\sinh(2\eta)}{\cosh(2\eta)-\cos(u)}.
\end{split}
\end{equation}
Expectation values of the charges in the thermodynamic limit are then
calculated as
\begin{equation}
\label{Fou}
  \frac{\vev{I_m}}{L}=\int_{-\pi}^\pi \frac{dp}{2\pi}\rho_r(p) i_m(p)=
\frac{1-q^{-2|m|}}{|m|}
\int_{-\pi}^\pi \frac{dp}{2\pi}\rho_r(p) e^{-imp}.
\end{equation}
It is a special property of this model that
the local charges measure the Fourier components of the root distribution.

\section{Quantum quenches and the Generalized Gibbs Ensemble in the
  $q$-boson model}

\label{sec:GGE}

We are interested in non-equilibrium situations in the $q$-boson
model. We assume that at $t=0$ the state is prepared in the initial state
\begin{equation*}
  \ket{\Psi(t=0)}=\ket{\Psi_0},
\end{equation*}
which is not an eigenstate of the Hamiltonian. It can be the ground
state of a local Hamiltonian, or any other state prepared according to
certain rules. Examples will be given in Section \ref{sec:120}.

The time evolution of physical observables is given by
\begin{equation*}
  \vev{\ordo(t)}=\bra{\Psi_0}e^{iHt}\ordo e^{-iHt}\ket{\Psi_0}.
\end{equation*}
We are interested in the large time behaviour of the observables in the
thermodynamic limit. Neglecting degeneracies in the spectrum the
long-time average in a finite volume case can be written as
\begin{equation}
\label{DE}
  \lim_{T\to\infty} \int_0^T dt \vev{\ordo(t)}=\sum_n
 |c_n|^2  \bra{n}\ordo\ket{n},\qquad c_n=\skalarszorzat{n}{\Psi_0}.
\end{equation}
The sum over eigenstates on the r.h.s. above can be interpreted
as a statistical physical ensemble, and it is called the 
Diagonal Ensemble (DE).
The weights of the DE are given by the squared overlaps with the
initial state, 
and it is an important question whether the predictions of the DE coincide with
those of a statistical physical ensemble, at least in the
thermodynamic limit. The system thermalizes, if the DE gives the same
mean values as a canonical or grand-canonical Gibbs Ensemble
(GE). This is the expected behaviour for a generic non-integrable
model. 

The situation is different in integrable models, where the existence
of the higher conserved charges prevents 
thermalization. It was suggested in \cite{rigol-gge} that in
these models the DE predictions should agree with those of a
Generalized Gibbs Ensemble (GGE) which includes all higher
charges.

In the case of the $q$-boson model the GGE density matrix can be defined as
\begin{equation}
\label{GGE}
  \rho_{GGE}=\frac{\exp\left(-\sum_{j=-\infty}^\infty \beta_jI_j\right)}
{\text{Tr} \exp\left(-\sum_{j=-\infty}^\infty \beta_jI_j\right)},
\end{equation}
where the Lagrange-multipliers are fixed by the requirement
\begin{equation}
\label{pI}
  \bra{\Psi_0}I_j\ket{\Psi_0}=\text{Tr}\left( \rho_{GGE} I_j\right).
\end{equation}
We also require $\beta_j=\beta_{-j}^*$, such that 
 $\rho_{GGE}$ is Hermitian. This condition is consistent with \eqref{pI}.

The GGE hypothesis states that for any local operator
\begin{equation*}
 \lim_{T\to\infty} \int_0^T dt \vev{\ordo(t)}  =\text{Tr}\left( \rho_{GGE} \ordo\right).
\end{equation*}
Time averaging is only required in finite volume, and it can be
omitted in the infinite volume limit.

The GGE was proven to be correct for free theories or models
equivalent to free fermions
\cite{rigol-gge,rigol-2,rigol-GETH,rigol-3,rigol-anyons,ising-quench-1,ising-quench-2,ising-quench-3,ising-quench-4,spyros-calabrese-gge},  
but counterexamples were found in \cite{JS-oTBA,sajat-oTBA}
in the case of the XXZ spin chain.
It is open question whether the GGE holds in other interacting
models, and what the precise conditions are for its validity.

In \cite{rigol-GETH} it was proposed that the GGE is valid whenever 
 the Generalized Eigenstate Thermalization Hypothesis
(GETH) holds. This hypothesis states that the mean values local operators in
the excited states only depend on the mean values of the charges. In
other words, if two excited states have mean values of the charges
that are close to each other, then all local correlations in the two
states will be close as well. 
Given that the initial state satisfies the
cluster decomposition principle, the Diagonal Ensemble \eqref{DE} 
 will be populated by states which have the same conserved charges as
 the initial state, such that the mean deviation for the densities of
 the charges becomes zero in the thermodynamic limit \cite{rigol-GETH,sajat-GETH}.
On the other hand, the GGE density matrix \eqref{GGE} produces states with the
 prescribed charges by definition. Therefore, if the GETH holds
 then the two ensembles 
 give the same results for the local operators, because the dominating states
 of both ensembles will have the same local correlations.
This is the reason why the validity of
 the GGE follows from the GETH \cite{rigol-GETH}.
We stress that
the weights $|c_n|^2$ of the DE need not be directly related to the generalized
 Boltzmann-weights of the GGE. 

\bigskip

We now argue that the GETH is valid in the
repulsive $q$-boson model. In this model there is only one
particle type in the spectrum, and the density of Bethe roots is
described by a single function $\rho_r(p)$. 
According to \eqref{Fou} the conserved charges
measure the Fourier components of $\rho_r(p)$ and
the root density can be reconstructed from the charges as
\begin{equation}
\label{rhoreconstr}
  \rho_r(p)=\frac{1}{L} \left(\vev{N}+\sum_{j=1}^\infty 
\frac{j(e^{ijp} \vev{I_j}+e^{-ijp} \vev{I_{-j}})}{1-q^{-2j}}
\right).
\end{equation}
The charges thus uniquely determine the root density, and
in order to prove the GETH we need to show that in the thermodynamic
limit the local correlators only depend on $\rho_r(p)$.
We are not able to prove this statement in full generality, but 
experience with other Bethe Ansatz solvable models 
suggests that it is in fact true. In subsection
\ref{sec:local} we prove it for a class of non-trivial local operators.

With this we have established that the GETH holds in the $q$-boson
model. As a consequence, the GGE should also hold for
any initial state satisfying the cluster decomposition principle.

Equation \eqref{rhoreconstr} shows that the charges uniquely determine
the root density and there is no need to obtain explicit expressions
for the Lagrange-multipliers entering the GGE density matrix. However,
for the sake completeness we show how to compute them.

The standard Thermodynamic Bethe Ansatz treatment of the GGE density
matrix \eqref{GGE} leads to the generalized TBA equations
\cite{caux-gge}:
\begin{equation}
\label{GGE-TBA}
 \eps(p)
=\beta_0+\sum_{m=1}^\infty \frac{1-q^{-2m}}{m}(\beta_m
e^{-imp}+\beta_m^* e^{imp}) -
\int_{-\pi}^\pi \frac{du}{2\pi} \varphi(p-u)\log(1+e^{-\eps(u)}),
\end{equation}
where $\eps(p)$ is the pseudoenergy defined as
\begin{equation}
  e^{\eps(p)}=\frac{\rho_h(p)}{\rho_r(p)}
\end{equation}
After the root density is obtained directly from \eqref{rhoreconstr}, the hole
density can be calculated from \eqref{rhoegy}. Substituting both
functions into the equation
\begin{equation}
\label{betak}
\begin{split}
&  \beta_0+\sum_{m=1}^\infty \frac{1-q^{-2m}}{m}(\beta_m
e^{-imp}+\beta_m^* e^{imp})=\\
&\hspace{3cm}\log\left(\frac{\rho_h(p)}{\rho_r(p)}\right)+
\int_{-\pi}^\pi \frac{du}{2\pi}
\varphi(p-u)\log\left(\frac{\rho_r(p)+\rho_h(p)}{\rho_h(p)}\right),
\end{split}
\end{equation}
the Lagrange-multipliers are obtained simply by Fourier
transformation. Note that all Lagrange-multipliers have a finite well-defined
value. 

We wish to remark that on the lattice all charges have finite
mean values, therefore the problem of infinities encountered for
interaction quenches in the continuum Bose gas
\cite{marci-ll-quench1,caux-stb-LL-BEC-quench} does not occur 
in the $q$-boson model. 
Also, equation \eqref{rhoreconstr} holds even
if there are logarithmic singularities in the root density,
because such functions are still members of $L^2([-\pi,\pi])$ and
therefore their Fourier series is well defined and converges almost
everywhere. Logarithmic singularities were encountered earlier in
other models \cite{caux-stb-LL-BEC-quench,JS-oTBA}, but in the
$q$-boson model they do not obstruct the validity of the GGE.

\bigskip

In the following subsection we show how to compute the GGE predictions
for a set of simple local observables. Specific quench problems are
considered in Section \ref{sec:120}.

\subsection{Local correlators in the GGE}

\label{sec:local}

For the $q$-boson model there are no results in the
literature for the excited state mean values of short range correlation functions. Here we
apply the Hellmann-Feynman theorem \cite{Hellmann-Feynman} to compute mean values of certain
local operators in states with arbitrary root density
$\rho_r(p)$. The GGE predictions are then calculated by substituting
the root density obtained from \eqref{rhoreconstr} into the results
presented below.
The method we apply was previously developed independently in
\cite{sajat-corr} and \cite{JS-oTBA}.

As a first example consider the Hermitian operator
\begin{equation*}
J_1=B_1 B^\dagger_2+B_1^\dagger B_2,
\end{equation*}
which is the operator density for $I_1+I_{-1}$. The finite volume mean
values in an arbitrary Bethe state are
\begin{equation}
\label{j1}
  \bra{\Psi}J_1\ket{\Psi}=\frac{1}{L} (1-q^{-2}) 
\sum_{j=1}^N  2\cos(p_j).
\end{equation}
We use the Hellmann-Feynman theorem to obtain the mean values of the
operator
\begin{equation*}
  J_1'=\frac{\partial}{\partial \eta}(B_1 B^\dagger_2+B_1^\dagger
  B_2)=
E_1 B^\dagger_2+E_1^\dagger B_2+B_1 E^\dagger_2+B_1^\dagger
  E_2,
\end{equation*}
where $E^\dagger=\partial B^\dagger/\partial \eta$ and $E=\partial B/\partial \eta$ are defined by their action on
the Fock states:
\begin{equation*}
  E_j\ket{n}_j=\sqrt{[n]'_q}\ket{n-1}_j\qquad
 E_j^\dagger\ket{n}_j=\sqrt{[n+1]'_q}\ket{n+1}_j,
\end{equation*}
where
\begin{equation*}
  [x]'_q=\frac{\partial}{\partial \eta}[x]_q=
\frac{2x q^{-2x} }{1-q^{-2}}+\frac{2(1-q^{-2x})q^{-2}}{(1-q^{-2})^2}.
\end{equation*}

Taking the derivative of \eqref{j1} with respect to 
 $\eta=\log(q)$ leads to
\begin{equation*}
\begin{split}
\bra{\Psi}J_1'\ket{\Psi}
=\frac{1}{L} 2q^{-2} 
\sum_{j=1}^N  2\cos(p_j)+
\frac{1}{L} (1-q^{-2}) \sum_{j=1}^N
(-2\sin(p_j)) \frac{dp_j}{d\eta}.
\end{split}
\end{equation*}
The derivatives $\frac{dp_j}{d\eta}$ can be obtained from the
logarithmic form of the Bethe equations:
\begin{equation}
\label{logBA}
  p_jl+\sum_{k\ne j} -i\log
  \frac{\sin((p_j-p_k)/2-i\eta)}{\sin((p_j-p_k)/2+i\eta)}=2\pi I_j.
\end{equation}
The quantum numbers $I_j\in\mathbb{Z}$ specify the state and can not
change as we vary $\eta$, therefore
\begin{equation}
\label{eltolas}
L \frac{dp_j}{d\eta}+\sum_{k\ne j}  \varphi(p_j-p_k)
 \left(\frac{dp_j}{d\eta}-\frac{dp_k}{d\eta}\right)+
\sum_{k\ne j} \tilde \varphi(p_j-p_k) 
=0,
\end{equation}
where $\varphi(p)$ is given by \eqref{vp} and
\begin{equation*}
  \tilde\varphi(p)=-\frac{2\sin(p)}{\cosh(2\eta)-\cos(p)}.
\end{equation*}
In a large volume $\frac{dp_j}{d\eta}\approx f(p_j)$, where $f(p)$ is
the so-called shift function. It follows from \eqref{eltolas} that it satisfies
\begin{equation*}
  f(p)+\int\frac{du}{2\pi}\varphi(p-u) \rho_r(u)(f(p)-f(u))+
\int \frac{du}{2\pi} \rho_r(u) \tilde\varphi(p-u)=0.
\end{equation*}
Using \eqref{rhoegy} we obtain
\begin{equation}
\label{fegy}
  f(p)(\rho_r(p)+\rho_h(p))-\int\frac{du}{2\pi}\varphi(p-u) \rho_r(u)f(u)+
\int \frac{du}{2\pi} \rho_r(u) \tilde\varphi(p-u)=0.
\end{equation}
This linear equation uniquely determines $f(p)$. 

Finally, the mean value of $J_1'$ is expressed as
\begin{equation}
\label{J1}
  \bra{\Psi}J_1'\ket{\Psi}=
4q^{-2}\int\frac{dp}{2\pi}\cos(p) \rho_r(p)
-2 (1-q^{-2})\int\frac{dp}{2\pi}\sin(p) f(p)\rho_r(p).
\end{equation}

Completely analogous results hold if we apply the Hellmann-Feynman
theorem to the remaining higher charges. Defining $J_m$ to be the
operator density of the Hermitian combination $I_m+I_{-m}$ we obtain
\begin{equation}
\label{Jm}
  \bra{\Psi}J_m'\ket{\Psi}=
4q^{-2m}\int\frac{dp}{2\pi}\cos(mp) \rho_r(p)
-2 \frac{1-q^{-2m}}{m}\int\frac{dp}{2\pi}\sin(mp) f(p)\rho_r(p),
\end{equation}
where $J_m'=\frac{\partial}{\partial \eta}J_m$ and
 the shift function is given by the solution of \eqref{fegy}. We
refrain from writing down the $J_m'$ in terms of local operators, as they
are easily obtained from the expressions of the charges $I_m$
\footnote{At present there are no closed form results known for $I_m$
  with arbitrary $m$. The cases $m=2$ and $m=3$ were computed in
  \cite{zvonarev-g3}, the $m=2$ case is given in
  eq. \eqref{charges}. Therefore, at present explicit expressions can be written
  down only
  for $J_1'$,  $J_2'$ and $J_3'$. Higher charges and higher $J'_m$  could be computed from
  the definition \eqref{Idef}. }. We just stress that all $J_m'$ are
non-trivial local operators which span at most $m+1$ sites.
 
\section{GGE for a class of initial states}

\label{sec:120}

In this section we derive the GGE solution for a special class of
initial states, which are given as tensor products of one-site
particle number eigenstates. Translationally invariant cases are
considered in \ref{sec:1}, whereas \ref{sec:20} deals with states that
break the translational invariance.

\subsection{Translationally invariant cases}

\label{sec:1}

We consider the states $\ket{F_n}$ in which there is exactly $n$
particle at each site:
\begin{equation*}
  \ket{F_n}=\otimes_{j=1}^L \ket{n}_j,\qquad n>0.
\end{equation*}
They are not eigenstates of the $q$-boson Hamiltonian. They can be
considered as ferromagnetic states pointing in a certain direction in
the infinite dimensional Fock space. Also, they are the ground states
of the infinitely repulsing Bose-Hubbard model at a given integer
filling.  The physically most relevant case
is $\ket{F_1}$ which is a state of uniform particle density 1. Quantum
quenches in the Bose-Hubbard model with initial state $\ket{F_1}$ were
studied in \cite{Bose-Hubbard-quench1}.

In the following we evaluate the $q$-boson GGE predictions for 
quenches starting from the 
$\ket{\Psi_0}=\ket{F_n}$. First we compute the expectation values of the
charges, then we reconstruct $\rho(p)$ from \eqref{rhoreconstr}, and
finally we give predictions for the local operators introduced in the
previous section.

It is easy to see that all charges $I_m$, $m>1$ are built from local
operators which have the form
\begin{equation}
\label{hukkl}
  (B_{j_1}^\dagger)^{n_1} \hdots (B_{j_2})^{n_2},
\end{equation}
such that $j_2>j_1$, $n_1,n_2>0$ and the dots stand for operators
acting on sites $j_k$ with $j_2>j_k>j_1$. In other words, the
leftmost and rightmost operators are ``unpaired'': 
a combination of the form
$B^\dagger_{j_k}B_{j_k}$ can only occur in the middle, but never on the two
ends of the operator product. This follows simply from the definition
of the transfer matrix \eqref{eq:T}: regarded as a power series in
$e^{-\lambda}$ the transfer matrix has terms of the form 
\begin{equation*}
  (B_{j_1}^\dagger) \hdots (B_{j_2}),
\end{equation*}
and after formally taking the logarithm only terms of the form
\eqref{hukkl} can arise.

As a consequence we obtain the remarkably simple result
\begin{equation*}
  \bra{F_n} I_m \ket{F_n}=0,
\end{equation*}
which follows from $_j\bra{n}(B_j)^k\ket{n}_j=0$ for arbitrary $n$, $k$ and
site $j$.

Applying \eqref{rhoreconstr} we find that 
in the quantum quench starting from the state $\ket{F_n}$ 
the resulting root density is constant and given
simply by the total particle density:
\begin{equation*}
  \rho_r(p)=n
\end{equation*}
The hole density can be calculated from \eqref{rhoegy}:
\begin{equation*}
  \rho_h(p)=1,
\end{equation*}
where we used $\int_{-\pi}^\pi \frac{dp}{2\pi}\varphi(p)=1$. The
filling fraction for these states is therefore
\begin{equation*}
  \frac{\rho_r(p)}{\rho_r(p)+\rho_h(p)}=\frac{n}{1+n}
\end{equation*}

A physical interpretation can be given as
follows. In the  initial states $\ket{F_n}$ all particles have
well-defined positions. The initial states are not eigenstates, so the
real-time dynamics of the system is non-trivial. Due to the initial
sharp localization in real space we can expect that the eigenstates
contributing to the dynamics
will be spread out maximally in momentum space. Finding a completely
constant root density in an interacting system is remarkable nevertheless, and it is a special
property of the system and the initial states chosen.

It is very easy to give predictions for the long-time limit of the
local operators $J'_m$ defined in the previous section. If $\rho_r(p)$
is constant then the unique solution of \eqref{fegy} is $f(p)=0$, and
from \eqref{Jm} we obtain
\begin{equation*}
\lim_{t\to \infty} \vev{J_m'(t)}=\vev{\rho_{GGE}J_m'}=0.
\end{equation*}
The operators $J_m'$ are special in the sense that they also have zero
mean value in the initial states:
\begin{equation*}
\vev{J_m'(0)}=  \bra{\Psi_0}J_m'\ket{\Psi_0}=0.
\end{equation*}
This follows from the fact that their structure is essentially the same as those of the
charges. However, they are not conserved in time. For example it can be
checked easily that 
\begin{equation*}
\left. \frac{d}{dt} \vev{J_1'(t)}\right|_{t=0}=i\bra{\Psi_0}[H,J_1']\ket{\Psi_0}\ne 0.
\end{equation*}
Therefore the prediction that all $\vev{J_m'(t)}$ approach zero in the
long-time limit is a highly non-trivial statement,
and it can be used as a test of the GGE.

For the sake of completeness we compute the Lagrange-multipliers for
this GGE. All relevant functions are constant and from
\eqref{betak} we obtain
\begin{equation}
\label{GGEbetak}
  \beta_0=\log \frac{1+n}{n},\qquad\text{and}\quad \beta_m=0
  \quad\text{for}\quad m\ne 0.
\end{equation}

To conclude this subsection we remark that any small departure from
the sharp localization of the one-site particle numbers changes the
resulting root densities. For example if the initial state is given by
\begin{equation*}
  \ket{\Psi_0}=\otimes_{j=1}^L \frac{\ket{1}_j+\alpha\ket{0}_j}{\sqrt{1+|\alpha|^2}},
\end{equation*}
then the mean values of the two simplest charges are
\begin{equation*}
\bra{\Psi_0}I_1\ket{\Psi_0}=\bra{\Psi_0}I_{-1}\ket{\Psi_0}=\chi^2 \frac{|\alpha|^2}{(1+|\alpha|^2)^2}.
\end{equation*}
According to \eqref{rhoreconstr} the root density $\rho(p)$ will have a non-vanishing
first Fourier component for any $\alpha\ne 0$.

\subsection{Breaking translational invariance}

\label{sec:20}

Let us define initial states which are not translationally invariant,
but still have fixed one-site particle numbers.
As examples we consider the states
\begin{equation}
\label{10def}
\begin{split}
 \ket{F_{10}}&= \otimes_{j=1}^{L/2}\
  \big(\ket{1}_{2j}\otimes \ket{0}_{2j-1}\big)\qquad\text{and}\qquad
  \ket{F_{20}}= \otimes_{j=1}^{L/2}\
  \big(\ket{2}_{2j}\otimes \ket{0}_{2j-1}\big).
\end{split}
\end{equation}
These states are invariant under translation by two sites.
It is an important question whether the full translational invariance gets
restored in the long-time limit. In Section \ref{sec:bosons} we
consider the free boson theory $(q=1)$ and demonstrate on a few simple
examples that mean values of local operators become translationally invariant indeed. In the cases
$q>1$ most of the hopping amplitudes in the
Hamiltonian are smaller than in the free case, but the particles can
still hop from any site to its neighbours irrespective of the
occupation number. Also, there is no one-site potential or 
any other term in the Hamiltonian which could ``freeze'' the
artificial order of the initial state. Therefore we conjecture that
translational invariance gets restored for any $q\ge 1$.

The GGE predictions for the steady state are derived easily. Mean
values of the charges $I_m$, $m>1$ are all zero due to the same
reasons as in the previous subsection. Therefore the
resulting root densities are constant and only depend on the average
particle number. For example $\rho(p)=1/2$ and $\rho(p)=1$ for
$\ket{F_{10}}$ and $\ket{F_{20}}$, respectively.

It follows from our considerations that the predictions of the GGE are
completely the same for any two initial states, if both are products of one-site
particle number eigenstates and the overall particle density is the
same. For example the initial states $\ket{F_{1}}$ and $\ket{F_{20}}$
should lead to the same long-time limit for any local quantity.
This is a surprising statement, and it can be
used as a  check of the GGE.

\section{Large $q$ limit: Equilibrium properties}

\label{sec:infty1}

In this section we treat the $q\to\infty$ limit of the model. We
review the special properties of this limiting case, and also establish new
results for a specific local operator: the Emptiness Formation
Probability. These results serve as a basis to study quantum quenches in
the large $q$ limit, which is considered in the next section.

In the $q\to\infty$ limit the local $q$-boson operators behave as
\begin{equation*}
  B_j\to \phi_j\qquad B^\dagger_j\to \phi^\dagger_j,
\end{equation*}
where the operators $\phi_j$, $\phi^\dagger_j$ are defined by their action 
\begin{equation*}
  \phi_j\ket{n}_j=\ket{n-1}_j\qquad \phi^\dagger_j \ket{n}_j=\ket{n+1}_j.
\end{equation*}
The Lax-operator is written as
\begin{equation*}
L(\lambda)=
\begin{pmatrix}
 e^\lambda & \phi^\dagger \\
\phi & e^{-\lambda} 
\end{pmatrix}.
\end{equation*}
and the Hamiltonian is
\begin{equation}
  \label{Hqq}
H=-\sum_{j=1}^L (\phi_j^\dagger \phi_{j+1}+ \phi_{j+1}^\dagger \phi_{j}-2N_j).
\end{equation}
This model attracted considerable attention, partly because it 
is closely related to the combinatorial problem of plane
partitions \cite{q-bozon-bog1,q-bozon-bog2,q-bozon-bog3,q-bozon-keiichi}.
In the literature it is often called the phase model.

The Algebraic Bethe Ansatz solution was first given in
\cite{q-bozon-izergin-kitanine-bog}, where equilibrium correlation
functions were computed as well. The coordinate space wave
functions were first computed in \cite{q-bozon-bog1}, where it was
shown that they are given by Schur polynomials. 
In our notations the coordinate Bethe Ansatz wave function can be
written as
\begin{equation}
\label{BetheStateQ}
 \ket{\{p\}_N}=\prod_{j=1}^N B(\lambda_j) \ket{0}=
\sum_{1\le x_1\le x_2\le \dots x_N\le L} 
 C_N(x_1,\dots,x_N)  
 \phi^\dagger_{x_1}\dots \phi^\dagger_{x_N}\ket{0},
\end{equation}
where the coefficients are
\begin{equation}
\label{coeff}
 C_N(x_1,\dots,x_N)=  
\frac{1}{\left(\prod_j e^{\lambda_j}\right)^{L+1}}
\frac{\det_N\Big( (a_j)^{k+x_k}\Big) }{\det_N \Big((a_j)^k\Big)},
\end{equation}
where
\begin{equation*}
  a_j=e^{2\lambda_j}=e^{ip_j}.
\end{equation*}
The Bethe equations take the following simple form:
\begin{equation}
\label{BeQ}
  (a_j)^{L+N} =(-1)^{N-1} \prod_{k=1}^N a_k.
\end{equation}
This can be obtained from the $q\to\infty$ limit of
\eqref{Be}, or directly from the ABA developed for the phase model
\cite{q-bozon-izergin-kitanine-bog}. 

If the rapidities satisfy the Bethe equations, then the 
norm of the
Bethe state \eqref{BetheStateQ} is
\begin{equation}
\label{normQ}
  \skalarszorzat{\{p\}_N}{\{p\}_N}
=L(L+N)^{(N-1)} \prod_{i<j} |a_i-a_j|^{-2}.
\end{equation}
This result was obtained in \cite{q-bozon-izergin-kitanine-bog} using
the Algebraic Bethe Ansatz, whereas in \cite{q-bozon-bog1} it was
shown that it follows from certain properties of the Schur
polynomials. The energy eigenvalues are given by
\begin{equation*}
  E_N=\sum_{j=1}^N e(p_j)\quad\text{where}\quad e(p)=4\sin^2(p/2),
\end{equation*}
whereas the eigenvalues for the higher charges are
\begin{equation}
\label{chargesinfty}
  \vev{I_m}=\frac{1}{|m|}\sum_{j=1}^N e^{-imp_j}.
\end{equation}
We also note that in the thermodynamic limit the relation between the
root and hole densities is simply
\begin{equation}
\label{rhoegyinfty}
  \rho_r(p)+\rho_h(p)=1+n,
\end{equation}
which follows from the $q\to\infty$ limit of \eqref{rhoegy} or
directly from the Bethe equations \eqref{BeQ}.

Results for correlation functions were also computed in
\cite{q-bozon-izergin-kitanine-bog} and
\cite{q-bozon-bog1,q-bozon-bog2,q-bozon-bog3}. Here we  consider
the $m$-site Emptiness Formation Probability (EFP), which is the probability
to have zero occupancy number on $m$ neighbouring sites. We derive new
formulas for the one-site and 
two-site EFP in arbitrary excited states. These results are used in the next section to give the
GGE predictions in the quench problems. 

Let us define the operators $\Pi_j$ which project to the zero-particle
state on site $j$. Their action is simply
\begin{equation*}
  \Pi_j\ket{n}_j=\delta_{n,0} \ket{0}_j.
\end{equation*}
The m-site EFP operator is given by
\begin{equation*}
  E^{(m)}=\prod_{j=1}^m \Pi_j.
\end{equation*}
In \cite{q-bozon-izergin-kitanine-bog} it was shown that the
normalized $m$-site EFP in a Bethe state is 
\begin{equation}
\label{ebbolindulj}
\bra{\{p\}_N}E^{(m)}\ket{\{p\}_N}
  =(1+n) \det Y^{(m)}
\end{equation}
with
\begin{equation*}
  Y^{(m)}_{jk}=\delta_{j,k}-\frac{1}{L}\frac{1}{1+n} \frac{\sin(\frac{m+1}{2}(p_j-p_k))}{\sin(\frac{1}{2}(p_j-p_k)}
\end{equation*}
and $n=N/L$. The thermodynamic limit of 
this expression is a Fredholm determinant. 

Here we consider the two simplest cases $m=1$ and $m=2$ and show 
that the determinant can be expressed using single sums over
the particles. In the thermodynamic limit we thus obtain
the EFP's as sums of products of simple integrals. This is a
huge simplification as opposed to the original result of a full Fredholm determinant.
For simplicity we only consider
states with zero total momentum, but this does not change the conclusions.

In the $m=1$ case we have
\begin{equation*}
  Y^{(1)}_{jk}=\delta_{j,k}-\frac{1}{L}\frac{1}{1+n} 
2\cos(\frac{1}{2}(p_j-p_k)).
\end{equation*}
Multiplying the $j$th row with $e^{ip_j/2}$ and the  $k$th column with  $e^{ip_k/2}$ leads to
\begin{equation*}
  \tilde  Y^{(1)}_{jk}=\delta_{j,k}a_j-F_{jk},\quad\text{with}\quad
F_{jk}=\frac{1}{L}\frac{1}{1+n} (a_j+a_k),
\end{equation*}
The matrix $F$ has rank 2,
therefore in the expansion of the determinant we only have terms where
at most 2 elements are chosen from $F$. This leads to
\begin{equation*}
\begin{split}
  \det \tilde  Y^{(1)}_{jk}&=
1-\sum_{j=1}^N \frac{1}{L}\frac{1}{1+n}  2
-\sum_{j<k} \frac{1}{a_ja_k} \left(\frac{1}{L}\frac{1}{1+n} \right)^2
(a_j-a_k)^2,
\end{split}
\end{equation*}
where we used $\prod_j a_j=1$. For the finite volume EFP we thus
obtain
\begin{equation*}
\bra{\{p\}_N}E^{(1)}\ket{\{p\}_N}
  =\frac{1}{1+n}-
\frac{1}{L^2}\frac{1}{1+n}\sum_{j,k} \frac{a_j}{a_k}.
\end{equation*}
Let us define renormalized higher charges as
\begin{equation}
\label{vagy1}
Q_m=|m|  \frac{\vev{I_m}}{L}=\frac{1}{L}\sum_{j=1}^N e^{-imp_j},
\end{equation}
where $I_m$ is given by \eqref{chargesinfty}.
In the thermodynamic limit we have
\begin{equation}
\label{vagy2}
  Q_m=\int_{-\pi}^\pi \frac{dp}{2\pi} \rho(p)e^{-imp}.
\end{equation}
Using this definition the one-site EFP is expressed simply as
\begin{equation}
\label{e1}
  \vev{E^{(1)}}=\frac{1}{1+n}\left(1- |Q_{1}|^2\right).
\end{equation}
Formula \eqref{e1} is valid both in finite volume and in the
thermodynamic limit. It is understood that
\eqref{vagy1} or \eqref{vagy2} has to be used depending on the situation.

We now calculate the two-site EFP.
As a first step we write the corresponding determinant as
\begin{equation*}
  \det  Y^{(2)}_{jk}= \prod_j a_j^{-2} \times  \det  \tilde Y^{(2)}_{jk},
\end{equation*}
where now
 \begin{equation*}
  \tilde  Y^{(2)}_{jk}=\delta_{j,k}a^2_j-F_{jk},\quad\text{with}\quad
F_{jk}=\frac{1}{L}\frac{1}{1+n} (a_j^2+a_ja_k+a^2_k),
\end{equation*} 
The matrix $F$ has at most rank 3, because it is a sum of three
matrices with rank 1. Therefore in the expansion of the determinant it
is enough keep terms where at most 3 elements are chosen from
$F$. This leads to
\begin{equation*}
\begin{split}
\det   Y^{(2)}_{jk}=&1-\sum_{j=1}^N \frac{1}{L}\frac{1}{1+n} 3+
\sum_{j<k} \frac{1}{L^2}\frac{1}{(1+n)^2} 
\left(6-\frac{a_j^2}{a_k^2}-\frac{a_k^2}{a_j^2}-2\frac{a_j}{a_k}-2\frac{a_k}{a_j}\right)
-\\
&-\sum_{j<k<l} \frac{1}{L^3}\frac{1}{(1+n)^3}
\left(\frac{a_j}{a_k}-\frac{a_k}{a_j}\right)^2
\left(\frac{a_j}{a_l}-\frac{a_l}{a_j}\right)^2
\left(\frac{a_k}{a_l}-\frac{a_l}{a_k}\right)^2.
\end{split}
\end{equation*}
After some tedious but elementary  calculations we obtain the EFP as
\begin{equation*}
\begin{split}
\bra{\{p\}_N}E^{(2)}\ket{\{p\}_N}=&
\frac{1}{(1+n)^2} 
-\frac{1}{L^2}\frac{2}{(1+n)^2}\sum_{j, k} 
\frac{a_j}{a_k}
-\frac{1}{L^2}\frac{1}{(1+n)^2}\sum_{j, k} 
\frac{a_j^2}{a_k^2}
\\
&
-\frac{1}{L^3}\frac{1}{(1+n)^2}\sum_{j,k,l} 
\left(\frac{a_ja_k}{a_l^2}+\frac{a_l^2}{a_ja_k}
\right).
\end{split}
\end{equation*}
Using the definition \eqref{vagy1} this can be expressed as
\begin{equation}
  \label{e2}
\begin{split}
\bra{\{p\}_N}E^{(2)}\ket{\{p\}_N}&=
\frac{1}{(1+n)^2} \left(
1-2|Q_1|^2-|Q_2|^2-((Q_{-1})^2Q_2+(Q_1)^2Q_{-2})
\right).
\end{split}
\end{equation}
This result remains valid in the thermodynamic limit if the definition
\eqref{vagy2} is used.

It is a special property of this system that the 1-site (or 2-site)
EFP could be expressed using the mean values of the first (or first
two) charges, respectively, and that the EFP does not depend on the
other details of the states. We now argue that this is a general
pattern: the $m$-site EFP only depends on the first $m$ charges and
the overall particle density. Also we show how to obtain the results
\eqref{e1} and \eqref{e2} directly in the thermodynamic limit.

In \cite{q-bozon-izergin-kitanine-bog} it was shown that the
thermodynamic limit of the formula \eqref{ebbolindulj} is the
Fredholm determinant
\begin{equation*}
 \vev{E^{(m)}}=(1+n)\det(1-\hat T^{(m)}), 
\end{equation*}
where $\hat T^{(m)}$ is an integral operator which acts on functions defined
on $[-\pi,\pi]$ as
\begin{equation*}
  (\hat T^{(m)}f)(p)=\int_{-\pi}^\pi \frac{dq}{2\pi}\ T^{(m)}(p,q) f(q),
\end{equation*}
where
\begin{equation*}
  T^{(m)}(p,q)=\frac{1}{1+n}\rho(y) K^{(m)}(p,q)\quad\text{with}\quad
K^{(m)}(p,q)=\frac{\sin(\frac{m+1}{2}(p-q))}{\sin(\frac{1}{2}(p-q))}.
\end{equation*}
The Fredholm determinant can be expressed as
\begin{equation}
\label{freddy}
  \det(1-\hat T^{(m)})=\sum_{k=0}^\infty
\frac{(-1)^k}{k!}\frac{1}{(1+n)^k} 
\left(\prod_{j=1}^k \int_{-\pi}^\pi dp_j\ \rho(p_j)\right) \det \Lambda^{(m)}_k,
\end{equation}
where $\Lambda^{(m)}_k$ is a $k$-by-$k$ matrix with elements given
by
\begin{equation}
\label{kilettfejtve}
  (\Lambda^{(m)}_k)_{ab}=K^{(m)}(p_a,p_b)=\frac{e^{i\frac{mp_a}{2}}}{e^{i\frac{mp_b}{2}}}+
\frac{e^{i\frac{(m-2)p_a}{2}}}{e^{i\frac{(m-2)p_b}{2}}}+\dots+ \frac{e^{i\frac{mp_b}{2}}}{e^{i\frac{mp_a}{2}}}.
\end{equation}
Note that $\Lambda^{(m)}_k$ is a sum of $m+1$ matrices
with rank 1, therefore its determinant is identically zero if
$k>m+1$. It is thus enough to keep the terms $k\le m+1$ in
\eqref{freddy}, which can be evaluated systematically using the
formula \eqref{kilettfejtve}. Note that for any $1\le a\le k$ the
variable $p_a$ only appears in the $a$th row or the $a$th column of
$\Lambda^{(m)}_k$. Therefore the highest power of $e^{ip_a}$ to appear
in the determinant is $e^{imp_a}$. It follows that the $m$-site EFP
only depends on the overall particle number and the charges $Q_l$ with $|l|\le m$.

\section{Large $q$ limit: Quantum Quenches}

\label{sec:infty}

In this section we investigate the quantum quenches in the $q\to\infty$ limit of the
system. In two cases we prove rigorously that
the GGE provides correct predictions for the stationary states. For
the quench from $\ket{\Psi_0}=\ket{F_{1}}$ we also compute the exact
time-dependence of the one-site EFP.

\subsection{Quantum quench from $\ket{\Psi_0}=\ket{F_{1}}$}

\label{hategy}

We consider the quench starting from the initial state
\begin{equation}
\label{inita}
  \ket{\Psi_0}=\ket{F_{1}}=\otimes_{j=1}^L \ket{1}_j,
\end{equation}
and evaluate the predictions of the Diagonal Ensemble \eqref{DE} in
the thermodynamic limit. We assume for simplicity that $L$ is even.

The overlaps with the initial state are non-zero only if $N=L$ and 
are given simply by a particular
component of the Bethe vector \eqref{BetheStateQ} with coordinates
$x_k=k$. Therefore, the normalized and squared overlaps are
\begin{equation}
\label{huhh}
\frac{| \skalarszorzat{F_1}{\{p\}_N}|^2}{
  \skalarszorzat{\{p\}_N}{\{p\}_N}}=
\frac{1}{N(2N)^{(N-1)}}
\left|\det \Big( (a_j)^{2k}\Big)\right|^2=
\frac{1}{N(2N)^{(N-1)}} \prod_{j<k} |a_j^2-a_k^2|^2.
\end{equation}
In deriving \eqref{huhh} we used that $|\prod_j a_j|=1$. 

The initial state is translationally invariant, therefore only states
with $\prod_j a_j=1$ can have a non-zero overlap. For these states the
Bethe equations \eqref{BeQ} are
\begin{equation*}
  a_j^{2L}=-1.
\end{equation*}
Solutions are given by
\begin{equation*}
  a_j=e^{i\frac{\pi (2I_j-1)}{2L}},\qquad I_j=1,2,\dots 2L.
\end{equation*}
The zero momentum Bethe states are thus given by the subsets
\begin{equation*}
  \{a\}_L \subset \{\omega\}_{2L}, \qquad \omega_k=e^{i\frac{\pi (2k-1)}{2L}}\qquad k=1,2,\dots 2L
\end{equation*}
satisfying the constraint $\prod_j a_j=1$.
The numbers $\omega_k$ can be paired such that
\begin{equation*}
  \{\omega\}_{2L}=\{(\omega_k,-\omega_k)\}_{k=1\dots L}.
\end{equation*}
It follows from formula \eqref{huhh} that
the overlap is non-vanishing only if exactly one rapidity is chosen from
each pair. In any other case at least one factor in \eqref{huhh} would
be zero.
Therefore, the states with non-vanishing overlap are given by
\begin{equation}
\label{ezek}
  a_j=s_j \omega_j,\quad\text{where}\quad s_j=\pm 1,\quad j=1\dots L,
\end{equation}
with the constraint that the total momentum is zero. We have
\begin{equation}
\label{zerom}
1=  \prod_{j=1}^N a_j=\prod_{j=1}^N (s_j \omega_j)=e^{i\pi N/2}\prod_{j=1}^N s_j.
\end{equation}
We assumed that $N$ is even, therefore the equation above can be
satisfied by choosing the first $N-1$ signs arbitrarily and then
fixing $s_N$ accordingly. It follows that there are a total number of $2^{N-1}$
states with non-vanishing overlap. 

The overlaps are functions of the variables $a_j^2$, therefore they
don't depend on the signs $s_j$. As a consequence, all non-vanishing
overlaps are equal and we obtain
\begin{equation}
\label{comp}
\frac{| \skalarszorzat{F_1}{\{p\}_N}|^2}{
  \skalarszorzat{\{p\}_N}{\{p\}_N}}
=
\frac{1}{2^{N-1}}.
\end{equation}
Comparing \eqref{comp} to \eqref{huhh} we obtain the identity
\begin{equation}
\label{bekellene}
 \prod_{1\le j<k\le N} \left|
e^{i\frac{2j\pi }{N}}-e^{i\frac{2k\pi}{N}}\right|^2= N^N.
\end{equation}
As a check of our calculations we prove this identity directly. The
l.h.s. can be written as
\begin{equation*}
 \prod_{1<j\ne k\le N} \left|
e^{i\frac{2j\pi }{N}}-e^{i\frac{2k\pi}{N}}\right|= 
\left(
\prod_{j=1}^{N-1} |1-e^{i\frac{2j\pi }{N}}|
\right)^N.
\end{equation*}
Therefore we need to show that
\begin{equation*}
  \prod_{j=1}^{N-1} |1-e^{i\frac{2j\pi }{N}}|=N.
\end{equation*}
Consider the polynomial
\begin{equation}
\label{zpol}
  \prod_{j=1}^{N-1} (z-e^{i\frac{2j\pi }{N}})=\frac{ \prod_{j=0}^{N-1}
    (z-e^{i\frac{2j\pi }{N}})}{z-1}=
\frac{z^N-1}{z-1}=1+z+z^2+\dots+z^{N-1}.
\end{equation}
In the second step we used that the product runs over all the roots of
the polynomial $z^N-1$. Substituting $z=1$ into \eqref{zpol} completes
the proof of \eqref{bekellene}.

\bigskip

With the solution \eqref{ezek} we have found the characterization of the states with non-vanishing
overlaps. These overlaps are all equal, therefore all
of these states have an equal weight in the Diagonal Ensemble. In the
thermodynamic limit the ensemble will be dominated by those states
where the signs $s_j$ are chosen randomly (without any particular
pattern depending on the rapidity), and this leads to a constant root
density in rapidity space:
\begin{equation*}
  \rho_r(p)=1.
\end{equation*}
This coincides with the prediction of the GGE for this particular
initial state. For the hole density we obtain $\rho_h(p)=1$ from
\eqref{rhoegyinfty}, and the filling fraction is $1/2$. 
This is in agreement with the description of the states in terms of
\eqref{ezek}.

We stress that even though the GGE
prediction for the root density has been 
confirmed, this does not mean that the Diagonal Ensemble is equal to
the GGE. The GGE density matrix produces all states with $\rho_r(p)=1$,
whereas the DE only includes those states which satisfy the constraint
$a_j^2\ne a_k^2$ for $j\ne k$. However, mean values of local operators only depend on
the root density, therefore the two ensembles lead to the same predictions.

\bigskip

The Emptiness Formation Probability is a physical observable with a
non-trivial time-dependence. In the inital state 
\begin{equation*}
  \bra{\Psi_0}E^{(m)}\ket{\Psi_0}=0, \quad \quad m=1,2,\dots
\end{equation*}
The GGE predictions for the long time limit can be calculated using
the results of the previous section. For the two simplest cases we
obtain from \eqref{e1} and \eqref{e2}
\begin{equation*}
  \lim_{t\to\infty}
  \bra{\Psi_0}E^{(1)}(t)\ket{\Psi_0}=\frac{1}{2}\quad
\text{and}\quad
 \lim_{t\to\infty} \bra{\Psi_0}E^{(2)}(t)\ket{\Psi_0}=\frac{1}{4}.
\end{equation*}

In this particular quench problem it is possible to go further and
compute the exact time dependence of these quantities. In the
following we derive an exact result for
$\bra{\Psi_0}E^{(1)}(t)\ket{\Psi_0}$. The computation of 2-site EFP is
left for further work.

\subsubsection{One-site EFP - Exact time evolution}

\label{jolkijott}

Form factors of $E_m$ were calculated in both
\cite{q-bozon-izergin-kitanine-bog} and \cite{q-bozon-bog1}. The
matrix element between two un-normalized off-shell Bethe states with arbitrary
rapidities $\ket{\{p^B\}_N}$ and  $\ket{\{p^C\}_N}$ reads
\begin{equation}
\label{FF}
\begin{split}
  \bra{\{p^C\}_N}E^{(m)}\ket{\{p^B\}_N}&=
\bra{0}\left(\prod_{j=1}^N C(p^C_j)\right)E^{(m)}\left(\prod_{j=1}^N B(p^B_j)\right)\ket{0}=\\
&=
\frac{\prod_{j=1}^L e^{i(ip^B_j+p^C_j)/2}}{\prod_{j<k}(e^{ip^B_j}-e^{ip^B_k})(e^{ip^C_k}-e^{ip^C_j})}
\det T^{(m)}(\{p^C\}_N,\{p^B\}_N),
\end{split}
\end{equation}
where
\begin{equation*}
  T^{(m)}_{jk}(\{p^C\}_N,\{p^B\}_N)=\frac{1}{e^{ip_k^C}-e^{ip_j^B}}
\Big(e^{i((2N+2L-1)p_k^C+p_j^B)/2}-
e^{i((2N+2L-2m-1)p_j^B+(2m+1)p_k^C)/2}
\Big).
\end{equation*}
If two rapidities coincide then the corresponding matrix element has
to be evaluated using the l'H\^opital rule. For example if $p_j^B\to
p_k^C$:
\begin{equation}
   T^{(m)}_{jk}\to  (N+L-m-1) e^{i(N+L-1)p_k^C}.
\end{equation}
If all rapidities coincide then we obtain the diagonal matrix
elements, which after normalization and using the Bethe equations lead to \eqref{ebbolindulj}.

In \eqref{FF} the state on the l.h.s. is a dual vector, but
it is not the complex conjugate of the ket vectors. In fact we have
the norm formula for on-shell states
\begin{equation}
  \label{norm2}
\bra{0}\prod_{j=1}^N C(p_j)\prod_{j=1}^N B(p_j)\ket{0}
=e^{iPN}L(L+N)^{(N-1)} \prod_{j\ne k} \frac{1}{e^{ip_k}-e^{ip_j}},
\end{equation}
where $P=\sum_j p_j$.

If both states are on-shell then 
\begin{equation*}
\det  T^{(1)}(\{p^C\}_N,\{p^B\}_N)=
(N+L)^N
\det \tilde T^{(1)}(\{p^C\}_N,\{p^B\}_N),
\end{equation*}
with
\begin{equation*}
  \tilde T^{(1)}(\{p^C\}_N,\{p^B\}_N)=F+G,
\end{equation*}
where
\begin{equation*}
  F_{jk}=-\frac{e^{ip_k^C}+e^{ip_j^B}}{N+L}.
\end{equation*}
and
\begin{equation*}
  G_{jk}=
  \begin{cases}
    0  &
    \text{if}\quad p_k^C\ne p_j^B\\
e^{-ip_k^C} &  \text{if}\quad p_k^C=p_j^B
  \end{cases}.  
\end{equation*}
We assumed here that $N$ is even. The matrix $F$ has rank 2, therefore
the determinant is non-vanishing 
only if the rank of $G$ is at least $N-2$. This means that $G$ must
have at least $N-2$ elements, therefore at least 
$N-2$ rapidities must coincide in the two states.

\bigskip

We consider the time evolution of the one-site EFP if the system is
quenched from the initial state \eqref{inita}. We parametrize the
rapidities of states with non-vanishing overlaps as
\begin{equation}
\label{ujpara}
  p_j=c_j+u_j\pi,\quad\text{where}\quad c_j=\frac{\pi (2j-1)}{2L}\quad\text{and}\quad u_j=0,1.
\end{equation}
The parameters $u_j$ can be chosen arbitrarily with the condition
\begin{equation}
\label{zerom2}
1=  \prod_{j=1}^N e^{ip_j}=e^{i\pi N/2}e^{i\pi\sum_{j=1}^N u_j}.
\end{equation}

The spectral expansion in a finite volume reads
\begin{equation}
  \label{wow2}
\begin{split}
&\bra{\Psi_0}E^{(1)}(t)\ket{\Psi_0}=
\sum_{\{u^B\}_N} \sum_{\{u^C\}_N} 
\frac{e^{it\sum_{j=1}^N (e(p_j^B)-e(p_j^C))}}{2^{N-2}} D \prod_{j=1}^N (e^{-ip_j^C/2} e^{-3ip_j^B/2} )
\det_N \tilde T^{(1)}(\{p^C\}_N,\{p^B\}_N),
\end{split}
\end{equation}
where the parametrization \eqref{ujpara} is implicit and $D$ is a sign
factor depending on the ordering of the rapidities:
\begin{equation*}
  D=\frac{\sqrt{\prod_{j\ne k} (e^{ip_j^B}-e^{ip_k^B})
      (e^{ip_k^C}-e^{ip_j^C})}}
{\prod_{j< k} (e^{ip_j^B}-e^{ip_k^B}) (e^{ip_k^C}-e^{ip_j^C})
}.
\end{equation*}
The determinant is non zero if there are at least $N-2$ coinciding
rapidity pairs: $p^B_j=p^C_k$ with some $j,k$.
Due to total momentum conservation there are only two possibilities:
Either the two states are exactly the same, or there are two
rapidities which differ. We treat these two cases separately. 

In the first case we have the diagonal elements
\begin{equation}
\label{finite-mean2}
\det_N \tilde T^{(1)}(\{p\}_N,\{p\}_N)=
-\frac{1}{4N^2} \sum_{j<k} \frac{(e^{ip_j}-e^{ip_k})^2}{e^{ip_j}e^{ip_k}}
=\frac{1}{2N^2}  \sum_{j<k} (1-\cos(p_j-p_k)).
\end{equation}
Here we used \eqref{zerom} again. The contribution of these terms in
\eqref{wow2} is
\begin{equation*}
  \sum_{\{u\}_N}\frac{1}{2^{N-1}} \frac{1}{N^2}  \sum_{j<k} (1-\cos(p_j-p_k)).
\end{equation*}
Using the parametrization \eqref{ujpara} we have
\begin{equation*}
  \sum_{\{u\}_N}\frac{1}{2^{N-1}} \frac{1}{N^2}  \sum_{j<k} (1-\cos(c_j-c_k+(u_j-u_k)\pi)),
\end{equation*}
where the sum runs over all sets of signs satisfying
\eqref{zerom2}. The sums can be exchanged and we obtain
\begin{equation*}
\frac{1}{2^{N-1}} \frac{1}{N^2}  \sum_{j<k}  \sum_{\{u\}_N}
(1-\cos(c_j-c_k+(u_j-u_k)\pi)) =\frac{1}{2^{N-1}} \frac{1}{N^2}
\sum_{j<k} 2^{N-1}=
\frac{N-1}{2N} .
\end{equation*}
We now consider the cases where there are two rapidity
differences. Let $a$ and $b$ with $a<b$ denote the positions of the differences. Then the
rapidities can be parametrized as
\begin{equation*}
  \begin{split}
    p_j^B =
    \begin{cases}
c_j+u_j\pi &\text{if}\quad j\ne a,b\\
   c_a+u_a\pi &\text{if}\quad j=a\\
   c_b+u_b\pi &\text{if}\quad j=b
    \end{cases}
\qquad\text{and}\qquad
    p_j^C =
    \begin{cases}
   c_j+u_j\pi  &\text{if}\quad j\ne a,b\\
   c_a+u_a\pi+\pi &\text{if}\quad j=a\\
   c_b+u_b\pi+\pi &\text{if}\quad j=b
    \end{cases},
  \end{split}
\end{equation*}
where condition \eqref{zerom2} is assumed.
The determinant is
\begin{equation}
\label{finite-mean3}
\det_N \tilde T^{(1)}(\{p^C\}_N,\{p^B\}_N)=-\frac{1}{2N^2}(1-\cos(c_a-c_b+(u_a-u_b)\pi)).
\end{equation}
The contribution of these cases to the EFP is
\begin{equation*}
  -\frac{1}{N^2}\sum_{a<b}  \sum_{\{u\}_{N}} \frac{1}{2^{N-1}}
e^{-4it(\cos(c_a+u_a\pi)+\cos(c_b+u_b\pi))}  
(1-\cos(c_a-c_b+(u_a-u_b)\pi)),
\end{equation*}
where we used that
\begin{equation*}
  e(p)-e(p+\pi)=-4\cos(p).
\end{equation*}
The summation over the $u$ variables can be performed leading to
\begin{equation*}
  -\frac{1}{N^2}\sum_{a<b}   
\left[
\cos(4\cos(c_a)t)\cos(4\cos(c_b)t)+
\sin(4\cos(c_a)t)\sin(4\cos(c_b)t)\cos(c_a-c_b)
\right].
\end{equation*}
Finally we obtain the finite-volume EFP as
\begin{equation*}
\bra{\Psi_0}E^{(1)}(t)\ket{\Psi_0}=
\frac{N-1}{2N}-
\frac{1}{N^2}\sum_{a<b}   
\left[
\cos(4\cos(c_a)t)\cos(4\cos(c_b)t)+
\sin(4\cos(c_a)t)\sin(4\cos(c_b)t)\cos(c_a-c_b)
\right].
\end{equation*}
Alternatively this can be written as
\begin{equation*}
\bra{\Psi_0}E^{(1)}(t)\ket{\Psi_0}=
\frac{1}{2}-
\frac{1}{2}\left(\frac{1}{N}\sum_{a}   
\cos(4\cos(c_a)t)\right)-\frac{1}{2}\left|\frac{1}{N}\sum_{a}
\sin(4\cos(c_a)t)e^{ic_a}\right|^2.
\end{equation*}

We can take the thermodynamic limit with a fixed $t$:
\begin{equation}
\label{jolettez}
\begin{split}
\bra{\Psi_0}E^{(1)}(t)\ket{\Psi_0}&=
\frac{1}{2}-\frac{1}{2}
\left(\int_{0}^\pi \frac{dp}{\pi} \cos(4\cos(p)t)\right)^2
-\frac{1}{2}
\left|\int_{0}^\pi \frac{dp}{\pi} \sin(4\cos(p)t)e^{ip}\right|^2.
\end{split}
\end{equation}
For the initial value this formula gives
$\bra{\Psi_0}E^{(1)}\ket{\Psi_0}=0$, as expected. In the long-time
limit the integrals become oscillatory and we have
\begin{equation}
\label{long-time1}
  \lim_{t\to\infty} \bra{\Psi_0}E^{(1)}(t)\ket{\Psi_0}=
\frac{1}{2}.
\end{equation}
This agrees with the GGE prediction. 

To our best knowledge formula \eqref{jolettez} is the first closed
form result 
result for the real time dynamics of a local observable in a genuinely
interacting infinite volume system.

\subsection{Quench from $\ket{\Psi_0}=\ket{F_{10}}$}

\label{hatketto}

Here we consider the quantum starting from the state $\ket{F_{10}}$
defined in \eqref{10def}. The initial state is not translationally
invariant, but we argued in the previous section
that translational invariance is restored in the long time
limit. Therefore we expect that the Diagonal Ensemble applies for the
local operators. 

The state $\ket{F_{10}}$ is invariant with respect to
translation by two sites, therefore the only states with non-vanishing
overlap are those with total pseudo-momentum equal to 0 or $\pi$. The
particle number is $N=L/2$.
The normalized overlaps are computed from \eqref{coeff} with
$x_k=2k$ and the norm formula \eqref{normQ}:
\begin{equation}
\label{huhh2}
\frac{| \skalarszorzat{F_{10}}{\{p\}_N}|^2}{
  \skalarszorzat{\{p\}_N}{\{p\}_N}}=
\frac{1}{L(L+N)^{(N-1)}}
\left|\det \Big( (a_j)^{3k}\Big)\right|^2=
\frac{1}{2N(3N)^{(N-1)}} \prod_{j<k} |a_j^3-a_k^3|^2.
\end{equation}
The Bethe equations take the form
\begin{equation*}
  a_j^{3N}=(-1)^{N-1} \prod_{k=1}^N a_k.
\end{equation*}
For simplicity we assume that $N$ is even and first consider states
with zero total momentum. In this case the solutions are of the form
\begin{equation*}
  a_j=e^{\frac{i\pi (2I_j-1)}{3N}},\qquad I_j=1\dots 3N.
\end{equation*}
Therefore, the Bethe state is described by a subset
\begin{equation*}
  \{a\}_N\subset \{w\}_{3N},\qquad w_j=e^{\frac{i\pi (2j-1)}{3N}}.
\end{equation*}
The numbers $w_j$ can be arranged in triplets as
\begin{equation*}
   \{w\}_{3N}=\{w_j,w_je^{i\pi/3},w_je^{2i\pi/3}\}_{j=1}^N.
\end{equation*}
It follows from \eqref{huhh2} that the overlap is non-vanishing
whenever exactly one member is chosen from each triplet. 
There are a
total number of $3^{N-1}$ such states.
The overlap does not depend on these choices,
and applying the identity \eqref{comp} in this case leads to 
\begin{equation}
\label{huhh3}
\frac{| \skalarszorzat{F_{10}}{\{p\}_N}|^2}{
  \skalarszorzat{\{p\}_N}{\{p\}_N}}=
\frac{1}{2\cdot 3^{N-1}}.
\end{equation}
It is easy to see that the same result is obtained for states with
total pseudo-momentum equal to $\pi$. Therefore all states with
non-vanishing overlap have an equal weight in the Diagonal
Ensemble. In the infinite volume limit the DE will be dominated by states where
the choices from the triplets are random, and this leads to
\begin{equation*}
  \rho_r(p)=\frac{1}{2},\qquad \rho_h(p)=1, \qquad  \frac{\rho_r(p)}{\rho_r(p)+\rho_h(p)}=1/3.
\end{equation*}
This result agrees with the
predictions of the GGE.

We also consider the long-time limit of the one-site and two-site EFP's. In the
initial state the one-site EFP is not translationally invariant:
it is equal to zero (one) on the even (odd) sites,
respectively. We conjectured that translational invariance is restored
in the long time limit, and from \eqref{e1} we obtain
\begin{equation*}
 \lim_{t\to \infty} \bra{\Psi_0}E^{(1)}(t)\ket{\Psi_0}=
\frac{2}{3}.
\end{equation*}
On the other hand, the two-site EFP is translationally invariant at
all times, its initial value is zero, and for the long time limit we
obtain from \eqref{e2}
\begin{equation*}
 \lim_{t\to \infty}  \bra{\Psi_0}E^{(2)}(t)\ket{\Psi_0}=
\frac{4}{9}.
\end{equation*}
The calculation of the exact time dependence is more challenging in
this case,
because the matrix elements between states with different total  momentum
have a more complicated structure than those treated in
\ref{jolkijott}. We leave this problem to further research.

\section{Free bosons}

\label{sec:bosons}

Here we consider the model in the $q\to 1$ limit, which is a free
bosonic theory defined by the Hamiltonian
\begin{equation}
  \label{Hq1}
H=-\sum_{j=0}^{L-1} (b_j^\dagger b_{j+1}+ b_{j+1}^\dagger b_{j}-2N_j).
\end{equation}
For later convenience we indexed the sites in \eqref{Hq1} from $0$ to $(L-1)$. 

It was shown in \cite{spyros-calabrese-gge} that if a model is
quenched from any state to a free Hamiltonian, then the GGE
holds whenever the initial state satisfies the cluster decomposition
principle. The proof of \cite{spyros-calabrese-gge} also applies to
the lattice model of free bosons, and the initial states considered in
the previous sections satisfy the cluster decomposition
principle. Therefore the GGE must be valid in these cases. 

The goal of the present section is to derive explicit formulas for
the time dependence of simple observables, and to demonstrate that the
GGE predictions are indeed correct, and that translational invariance
is restored in the large time limit. We consider two initial states:
\begin{equation*}
  \ket{\Psi_0}=\ket{F_1} \quad\text{and}\quad \ket{\Psi_0}=\ket{F_{20}}.
\end{equation*}
They have the same particle density $n=1$ and according to the GGE
they should lead to the same stationary state. 
As physical observables
we choose the one-site particle number 
operators $N_j$ and their square $N_j^2$. We first evaluate their
exact time evolution, and then show that in the long time limit they
approach the GGE predictions. The calculations below are
straightforward and elementary. Nevertheless, we felt that it is useful to
present them, so that both the  $q\to\infty$ and $q=1$ points
can be benchmarks for the generic $q$ case.

The model is diagonalized with the Fourier modes of the one-site
bosonic operators:
\begin{equation*}
  \tilde b_k=\frac{1}{\sqrt{L}}\sum_{n=0}^{L-1} b_n e^{-i2\pi kn/L}\qquad
  \qquad\qquad
  \tilde b^\dagger_k=\frac{1}{\sqrt{L}}\sum_{n=0}^{L-1} b^\dagger_n e^{-i2\pi kn/L}.
\end{equation*}
The following commutation relation holds:
\begin{equation*}
  [\tilde b_j,\tilde b_k^\dagger]=\delta_{j,k}.
\end{equation*}
In terms of the Fourier modes the Hamiltonian can be written as
\begin{equation*}
H=\sum_{k=0}^{L-1} \epsilon_k  \tilde b_k^\dagger \tilde b_k,
\end{equation*}
where
\begin{equation*}
  \epsilon_k=4\sin^2(p_k/2),\quad\text{with}\quad p_k=\frac{2\pi k}{L}.
\end{equation*}
Therefore the time dependence of the operators (in the Heisenberg picture) is
\begin{equation*}
  \tilde b^\dagger_k(t)=  \tilde b^\dagger_k e^{-i \epsilon_k t}\qquad
  \tilde b_k(t)=  \tilde b_k e^{i \epsilon_k t}.
\end{equation*}
We will consider the particle number operators on site 0 and 1:
\begin{equation*}
  N_0=\frac{1}{L} \sum_{j,k} \tilde b_j^\dagger \tilde b_k\qquad
  N_1=\frac{1}{L} \sum_{j,k}  \tilde b_j^\dagger \tilde b_k  e^{2\pi i (j-k)/L}.
\end{equation*}
Their time-dependence is
\begin{equation*}
 N_0(t)=\frac{1}{L} \sum_{j,k} \tilde b_j^\dagger \tilde b_k   e^{i
   (\epsilon_j-\epsilon_k) t}\qquad
  N_1(t)=\frac{1}{L} \sum_{j,k} \tilde
  b_j^\dagger \tilde b_k  e^{2\pi i (j-k)/L} e^{i
   (\epsilon_j-\epsilon_k) t}.
\end{equation*}
For their squares we obtain
\begin{equation}
\label{negyzetesek}
\begin{split}
  N_0^2&=\frac{1}{L^2} \sum_{j_1,j_2,k_1,k_2} 
\tilde b_{j_1}^\dagger \tilde b_{j_2}^\dagger \tilde b_{k_1}\tilde
b_{k_2}+N_0\\
  N_1^2&=\frac{1}{L^2} \sum_{j_1,j_2,k_1,k_2} 
\tilde b_{j_1}^\dagger \tilde b_{j_2}^\dagger \tilde b_{k_1}\tilde
b_{k_2}  e^{2\pi i (j_1+j_2-k_1-k_2)/L} +N_1,
\end{split}
\end{equation}
with the time dependence given by
\begin{equation*}
\begin{split}
  N_0^2(t)&=\frac{1}{L^2} \sum_{j_1,j_2,k_1,k_2} 
\tilde b_{j_1}^\dagger \tilde b_{j_2}^\dagger \tilde b_{k_1}\tilde
b_{k_2}
e^{i (\epsilon_{j_1}+\epsilon_{j_2}-\epsilon_{k_1}-\epsilon_{k_2})
  t}+N_0(t)\\
  N_1^2(t)&=\frac{1}{L^2} \sum_{j_1,j_2,k_1,k_2} 
\tilde b_{j_1}^\dagger \tilde b_{j_2}^\dagger \tilde b_{k_1}\tilde
b_{k_2} e^{2\pi i (j_1+j_2-k_1-k_2)/L}
e^{i (\epsilon_{j_1}+\epsilon_{j_2}-\epsilon_{k_1}-\epsilon_{k_2})
  t}+N_1(t).
\end{split}
\end{equation*}

\subsection{Quench from $\ket{\Psi_0}=\ket{F_1}$}

In this case the initial state is given by
\begin{equation*}
  \ket{\Psi_0}=\ket{F_1}=b^\dagger_{l-1} \dots b^\dagger_1 b_0^\dagger \ket{0}.
\end{equation*}
This state is translationally invariant, therefore it is enough to
consider the operators on site 0. In calculating the local observables
below we will make use of 
 the commutation relations
\begin{equation*}
  [\tilde b_k,b_n^\dagger]=\frac{1}{L} e^{-i2\pi kn/L},\qquad
  [b_n,\tilde b_k^\dagger]=\frac{1}{L} e^{i2\pi kn/L}.
\end{equation*}
For the time dependence of the particle number operator we obtain
\begin{equation*}
\begin{split}
  \bra{F_1}N_0(t)\ket{F_1}&=\frac{1}{L^2}\sum_{j,k}  e^{i
    (\epsilon_j-\epsilon_k) t} \sum_n e^{-i2\pi (k-j)n/L}=\frac{1}{L} \sum_j 1=1.
\end{split}
\end{equation*}
This is the expected result, because the total particle number is
conserved and the system is translationally invariant at all times.

For the expectation value of $N_0^2$ we obtain
\begin{equation*}
\begin{split}
  \bra{F_1}N_0^2(t)\ket{F_1}&=\bra{F_1}N_0(t)\ket{F_1}+\frac{1}{L^4} \sum_{j_1,j_2,k_1,k_2} 
e^{i (\epsilon_{j_1}+\epsilon_{j_2}-\epsilon_{k_1}-\epsilon_{k_2})  t}
\sum_{n_1> n_2} ( e^{-i2\pi ((k_1-j_1)n_1+(k_2-j_2)n_2)/l}+\text{perm.})\\
&=3-\frac{2}{L^3} \sum_{j_1,j_2,k_1} 
e^{i
  (\epsilon_{j_1}+\epsilon_{j_2}-\epsilon_{k_1}-\epsilon_{j_1+j_2-k_2})
  t}.
\end{split}
\end{equation*}
Performing the infinite volume limit for fixed $t$ leads to
\begin{equation*}
\bra{F_1}N_0^2(t)\ket{F_1}=3-
2\int_{-\pi}^\pi \frac{dp_1}{2\pi} \frac{dp_2}{2\pi} \frac{dq_1}{2\pi}
e^{i (\epsilon(p_1)+\epsilon(p_2)-\epsilon(q_1)-\epsilon(p_1+p_2-q_1)) t},
\end{equation*}
where
\begin{equation*}
  \epsilon(p)=4\sin^2(p/2).
\end{equation*}
For $t=0$ the above formula yields $\bra{F_1}N_0^2\ket{F_1}=1$, as
expected. For $t\to\infty$ the integrals become strongly oscillatory 
and this leads to 
\begin{equation*}
  \lim_{t\to \infty}   \bra{F_1}N_0^2(t)\ket{F_1}=3.
\end{equation*}

\subsection{Quench from $\ket{\Psi_0}=\ket{F_{20}}$}

Here we consider the quench from the initial state
\begin{equation*}
  \ket{\Psi_0}=\ket{F_{20}}=
\frac{1}{\sqrt{2^{L/2}}}
b^\dagger_{l-2}b^\dagger_{l-2} \dots b^\dagger_2 b_2^\dagger  b^\dagger_0 b_0^\dagger \ket{0}.
\end{equation*}
We compute the time evolution of the one-site particle number
operators. The initial state is two-site shift invariant,  therefore it
is sufficient to consider the operators $N_0(t)$ and $N_1(t)$.

For the time evolution of $N_0(t)$ we obtain
\begin{equation*}
\begin{split}
  \bra{F_{20}}N_0(t)\ket{F_{20}}&=\frac{1}{2L^2}\sum_{j,k}  e^{i
    (\epsilon_j-\epsilon_k) t} \sum_{n=0}^{(l-2)/2}4 e^{-i2\pi (k-j)2n/L}=\\
&=\frac{1}{L} \sum_{j,k}  e^{i  (\epsilon_j-\epsilon_k) t}
(\delta_{j,k}+\delta_{j-k,L/2})=\\
&=1+\frac{1}{L} \sum_{j} e^{i  (\epsilon_j-\epsilon_{j+L/2}) t}.
\end{split}
\end{equation*}
In the thermodynamic limit this leads to
\begin{equation*}
  \bra{F_{20}}N_0(t)\ket{F_{20}}=1+\int_{-\pi}^\pi \frac{dp}{2\pi} e^{i  (\epsilon(p)-\epsilon(p+\pi)) t}.
\end{equation*}
Similarly 
\begin{equation*}
\begin{split}
  \bra{F_{20}}N_1(t)\ket{F_{20}}&=\frac{1}{2L^2}\sum_{j,k}  e^{i
    (\epsilon_j-\epsilon_k) t} 
e^{2\pi i (j-k)/L}
\sum_{n=0}^{(l-2)/2}4 e^{-i2\pi (k-j)2n/L}=\\
&=\frac{1}{L} \sum_{j,k}  e^{i  (\epsilon_j-\epsilon_k) t}
(\delta_{j,k}-\delta_{j-k,L/2})=\\
&=1-\frac{1}{L} \sum_{j} e^{i  (\epsilon_j-\epsilon_{j+L/2}) t},
\end{split}
\end{equation*}
leading to
\begin{equation*}
  \bra{F_{20}}N_1(t)\ket{F_{20}}=1-\int_{-\pi}^\pi \frac{dp}{2\pi} e^{i  (\epsilon(p)-\epsilon(p+\pi)) t}.
\end{equation*}
In the initial state $\bra{F_{20}}N_0\ket{F_{20}}=2$ and
$\bra{F_{20}}N_1\ket{F_{20}}=0$, whereas in the long time limit we have
\begin{equation*}
  \lim_{t\to\infty} \bra{F_{20}}N_0(t)\ket{F_{20}}=
 \lim_{t\to\infty}\bra{F_{20}}N_1(t)\ket{F_{20}}=1.
\end{equation*}
For the squared operators similar but somewhat lengthier calculations
result in
\begin{equation*}
\begin{split}
 \bra{F_{20}}N_0^2(t)\ket{F_{20}}=&3+
5\int_{-\pi}^\pi \frac{dp}{2\pi} e^{i  (\epsilon(p)-\epsilon(p+\pi))
  t}+2\left(
\int_{-\pi}^\pi \frac{dp}{2\pi} e^{i  (\epsilon(p)-\epsilon(p+\pi)) t}\right)^2
+\\
&-3\int_{-\pi}^\pi \frac{dp_1}{2\pi} \frac{dp_2}{2\pi} \frac{dq_1}{2\pi}
e^{i (\epsilon(p_1)+\epsilon(p_2)-\epsilon(q_1)-\epsilon(p_1+p_2-q_1))
  t}\\
&-3\int_{-\pi}^\pi \frac{dp_1}{2\pi} \frac{dp_2}{2\pi} \frac{dq_1}{2\pi}
e^{i (\epsilon(p_1)+\epsilon(p_2)-\epsilon(q_1)-\epsilon(p_1+p_2-q_1+\pi))
  t}
\end{split}
\end{equation*}
and
\begin{equation*}
\begin{split}
 \bra{F_{20}}N_1^2(t)\ket{F_{20}}=&3-
5\int_{-\pi}^\pi \frac{dp}{2\pi} e^{i  (\epsilon(p)-\epsilon(p+\pi))
  t}+2\left(
\int_{-\pi}^\pi \frac{dp}{2\pi} e^{i  (\epsilon(p)-\epsilon(p+\pi)) t}\right)^2
+\\
&-3\int_{-\pi}^\pi \frac{dp_1}{2\pi} \frac{dp_2}{2\pi} \frac{dq_1}{2\pi}
e^{i (\epsilon(p_1)+\epsilon(p_2)-\epsilon(q_1)-\epsilon(p_1+p_2-q_1))
  t}\\
&+3\int_{-\pi}^\pi \frac{dp_1}{2\pi} \frac{dp_2}{2\pi} \frac{dq_1}{2\pi}
e^{i (\epsilon(p_1)+\epsilon(p_2)-\epsilon(q_1)-\epsilon(p_1+p_2-q_1+\pi))
  t}.
\end{split}
\end{equation*}
At $t=0$ we have  $\bra{F_{20}}N_0^2\ket{F_{20}}=4$ and
$\bra{F_{20}}N_1^2\ket{F_{20}}=0$, whereas in the long time limit 
\begin{equation*}
  \lim_{t\to\infty} \bra{F_{20}}N_0^2(t)\ket{F_{20}}=
 \lim_{t\to\infty}\bra{F_{20}}N_1^2(t)\ket{F_{20}}=3.
\end{equation*}
Thus we have demonstrated on these simple examples that non-trivial
observables indeed become translationally invariant, and approach the
same values as in the case of the initial state $\ket{F_{1}}$. 

\subsection{GGE predictions}

In a free theory the GGE can be built conveniently using the mode occupation
numbers. We define
\begin{equation*}
  \rho_{GGE}=\frac{e^{-\sum_{j=0}^L \beta_j \tilde I_j}}
{\text{Tr}( e^{-\sum_{j=0}^L \beta_j \tilde I_j})},\qquad
\tilde  I_j=\tilde b^\dagger_j \tilde  b_j.
\end{equation*}
This construction is equivalent to a GGE built from the local charges,
which are the Fourier components of $\tilde I_j$:
\begin{equation*}
  I_l=\sum_{j=0}^{L-1} e^{i\frac{2\pi jl}{L}}\tilde I_j=
\sum_{n=0}^{L-1} b^\dagger_{n+l} b_n.
\end{equation*}
The Lagrange multipliers have to be fixed by the initial values of the
charges, which are
\begin{equation*}
  \bra{F_1}\tilde I_j\ket{F_1}=
  \bra{F_{20}}\tilde I_j\ket{F_{20}}=1.
\end{equation*}
On the other hand
\begin{equation*}
  \text{Tr}\left(\rho_{GGE} \tilde I_j\right)=
 \frac{\text{Tr}\Big( e^{-\beta_j  \tilde I_j }\tilde I_j\Big)
}
{\text{Tr}\Big( e^{-\beta_j \tilde I_j}\Big) }=
\frac{1}{e^{\beta_j}-1},
\end{equation*}
which leads to $\beta_j=\log(2)$. All Lagrange multipliers are
equal, therefore the GGE density matrix only depends on the total
particle number operator. Note that the same result was found also in
the interacting case, see eq. \eqref{GGEbetak}. 

The GGE predictions for the operator $N_0^2$ can be evaluated using
\eqref{negyzetesek} and Wick's theorem and we find
\begin{equation*}
  \text{Tr} \left(\rho_{GGE}N_0^2\right)=3.
\end{equation*}
This result agrees with the asymptotic values derived in the previous
two subsections, as expected.

\section{Discussion and Outlook}

\label{sec:vege}

In this paper we considered quantum quenches in the $q$-boson
model. First we showed that the Generalized Eigenstate Thermalization
Hypothesis holds in this system, therefore the GGE gives the
correct asymptotic states if the initial state satisfies the cluster
decomposition principle. The role of the latter is to ensure that the
Diagonal Ensemble only includes states having the same mean values
for the charges as the initial state.

Concentrating on simple initial states which
have fixed one-site occupation numbers we were able to provide the GGE
predictions. Surprisingly we found that for these initial states the
resulting root densities are $\rho_r(p)=n$, where $n$
is the overall particle
density. This result means that any two states within this
family  which have the same
particle density will also have the same stationary behaviour.

We also considered the $q\to\infty$ limit of the system,
where the exact overlaps are given by Schur polynomials, which can be
expressed as determinants. For two initial states we were able to
determine which states populate the Diagonal Ensemble. As a
consequence
we proved that in the thermodynamic limit the GGE predictions are correct. We believe
that this is the first time that the GGE was proven to be valid in a
model which is neither a free theory, nor solvable by free
fermions. For the quench starting from $\ket{\Psi_0}=\ket{F_1}$ we
calculated the exact time dependence of the one-site Emptiness Formation
Probability (EFP). To our best knowledge this is the first time that an
analytic result has been obtained for the real time dynamics of an
observable in a genuinely interacting system, valid both in finite
volume and in the thermodynamic limit.

\bigskip

A few comments about our results are in order.

The initial states that we considered are very
special. They are not entangled at all, and they are pure Fock states in the
local bosonic basis. However, they are very reasonable choices from a
physical point of view. The states $\ket{F_n}$ are exact ground states of the infinitely
repulsing Bose-Hubbard model with a given filling $n$, or they are
approximate ground states for large repulsion parameters. Also, they
could be realized in experimental situations. The state
$\ket{F_{10}}$ can be considered as an analogue of the N\'eel state in the
spin-$1/2$ XXZ spin chain,  which has been the subject of a large
body of recent theoretical works (see \cite{sajat-oTBA,JS-oTBA,Neel-quench-demler-gritsev} and
references therein). The XXZ model can be described as a theory
of interacting fermions with attractive coupling, whereas here we dealt with an interacting
bosonic system with repulsive coupling, so the physical behaviour can be markedly
different. However, our initial states themselves are not more
exceptional than the N\'eel state or other states considered elsewhere in the literature. 

The $q$-boson model has been used previously as a lattice
regularization of the continuum Lieb-Liniger model
\cite{zvonarev-g3,marci-ll-quench1}. Here we focused on the lattice
model in its own right.
 In fact, all  initial states become ill-defined if we take the
 scaling limit towards the Lieb-Liniger model, because they
 have fixed particle density in lattice units, which leads to
 infinite density in the continuum limit. One could consider a special
 limit where the initial state is also changed as we approach
 the Lieb-Liniger model, such that the final particle density is
 finite. This way we would obtain an initial state where particles are
 localized with Dirac-delta functions at a fixed distance from each
 other. However, these states are not normalizable and they excite Bethe
 states with arbitrarily high energy, therefore this would not be a
 well-defined quench problem either. We wish to stress that our
 initial states have just the opposite structure as those considered
 previously in \cite{marci-ll-quench1,caux-stb-LL-BEC-quench}: they are sharply localised in
 real space, as opposed to the Bose-Einstein condensate (BEC) states
 which are localised in momentum space.

In Section \ref{sec:infty} we considered the $q\to\infty$ limit and
in two cases we obtained the exact overlaps in a product form (eqs. \eqref{huhh}
and \eqref{huhh2}). The
simplicity of this result is certainly a special property of both
the model and the initial states, but already these cases show 
different behaviour than expected.
In
quantum quenches of the XXZ spin 
chain and the Lieb-Liniger model
 it was found that in the on-shell case (when the rapidities satisfy the Bethe
 equations) only those states have non-zero overlap which are
 formed out of $(p,-p)$ rapidity pairs
\cite{caux-stb-LL-BEC-quench,Caux-Neel-overlap1,Caux-Neel-overlap2}.
In the $q$-boson model, on the other hand, we found completely
different conditions. For example for $\ket{\Psi_0}=\ket{F_1}$ the
requirement for the overlap to be non-vanishing is
\begin{equation*}
  e^{2ip_j}\ne e^{2ip_k}\quad\text{for}\quad j\ne k.
\end{equation*}
The
$q\to\infty$ limit of the model is regular, the finite volume overlaps are
continuous functions of $q$, therefore the pairing requirement does not hold
for finite $q$ either. 

In the previously considered quench problems
the Quench Action method \cite{quench-action} was used to find
the stationary states
\cite{caux-stb-LL-BEC-quench,JS-oTBA,sajat-oTBA}. The Quench Action is
a functional of the root and hole densities and is given by a combination 
of the overlaps and the micro-canonical 
entropy associated to a given root configuration. Finding the minimum of the
QA provides the states which populate the system
after the quench.
In the XXZ chain and the Lieb-Liniger model
the on-shell overlaps could be
written in the form \cite{caux-stb-LL-BEC-quench,Caux-Neel-overlap1,Caux-Neel-overlap2}
\begin{equation}
\label{joforma}
  \frac{| \skalarszorzat{\Psi_0}{\{p\}_N}|^2}{
  \skalarszorzat{\{p\}_N}{\{p\}_N}}=C(\{p\}_N)\prod_{j=1}^N v(p_j),
\end{equation}
where the pre-factor $C$ behaves as $\ordo(L^0)$ in the thermodynamic limit.
This made it possible to write the QA  as simple integrals over the root and hole
densities, thus making a TBA-like analysis possible. 
In the present case the on-shell overlap formulas \eqref{comp} and
\eqref{huhh3} are so
simple that the conclusions could be drawn immediately and there was
no need to apply the machinery of the QA approach. However, the
 non-zero on-shell overlaps are actually of the form
\eqref{joforma}. For example for $\ket{\Psi_0}=\ket{F_1}$ we have
\begin{equation*}
  v(p)=\frac{1}{2}\quad\text{and}\quad C=2
\end{equation*}
This trivially leads to a constant root density.

As a final comment we note that even though we confirmed the GGE
hypothesis in the $q$-boson model, we did not make use of the GGE density
matrix in any way. We assumed that the Diagonal Ensemble is valid, and
argued that the GETH holds. Then the predictions for the long-time limit of
the observables could be calculated as soon as the root density
$\rho_r(p)$ was obtained from the charges. The generalized TBA
equations \eqref{GGE-TBA} were only used to calculate the Lagrange
multipliers. Therefore it might be more appropriate to call our
results the ``DE+GETH predictions'' instead of the ``predictions of
the GGE''.
This behaviour might be a generic feature of interacting integrable models: If
the GETH holds, then there must be a one-to-one correspondence between
the charges and the root densities, therefore the GGE density matrix and
the associated TBA equations are not
needed. On the other hand, if the GETH does not hold, then the TBA
analysis of the GGE density matrix is expected to give wrong
predictions \cite{sajat-GETH,andrei-gge}. 

\bigskip

Below we list a number of open questions, which deserve further investigation.

\begin{itemize}
\item For a generic $q$
is there a formula of the form \eqref{joforma} for the overlaps?
The wave functions are given by Hall-Littlewood functions
\cite{q-bozon-coo-BA}, and the overlaps with the pure Fock states
  are specific components
of the wave function. It might be possible that determinant formulas
could be found, if one makes use of the Bethe equations. Note that the
Quench Action method can only agree with the GGE if the function
$v(p)$ appearing in \eqref{joforma} is a constant. In any other
case the resulting root density would not be constant, and this would
contradict the GGE prediction.

\item  Results for local correlation functions of the $q$-boson model
 are very limited. Mean values are only known for the charges and
 their derivatives with respect to $\eta=\log(q)$ (see Subsection
 \ref{sec:local}). It would be useful 
 to derive new results for other operators, for example the $m$-site
 Emptiness Formation Probability.
\item If the initial state is not translationally invariant, does this
  symmetry get restored in the long time limit? We argued that it
  does, but were not able to give a proof.
\item Is it possible to derive GGE predictions for other initial
  states? There are no closed form formulas for the higher
  charges, but a truncated GGE can be established with the already
  available results   \cite{marci-ll-quench1}.
In the case of the XXZ spin chain a generating function for the higher
charges was constructed in \cite{essler-xxz-gge} and it was shown how
to compute it for simple product states.
It is an open question whether the method of \cite{essler-xxz-gge} can be
generalized to the $q$-boson model.
\item In the $q\to\infty$ limit both the overlaps and the matrix
  elements of local operators are known and they take a relatively
  simple form
  \cite{q-bozon-izergin-kitanine-bog,q-bozon-bog1,q-bozon-bog3}. Moreover,
  the Bethe equations can be solved explicitly. These two properties
  allowed us to compute the time-dependence of the
  one-site EFP in the case of $\ket{\Psi_0}=\ket{F_1}$. We believe that this
  result could be extended to other operators. For example, the
  $m$-site EFP could be calculated with the same methods, such that
  result could be expressed as sums of products of simple integrals. 
 It is an open question, whether exact calculations are possible for other
operators or other initial states.
\item In Section \ref{sec:infty1} we found that in the $q\to\infty$
  limit the excited state mean values of the $m$-site EFP only depend
  on the mean values of the first $m$ charges and the particle
  density. This opens up a way to give exact GGE predictions even for
  those initial states where the full root density can not be
  reconstructed. For example, if the mean value of the first charge
    $I_1=\sum_j \phi_j^\dagger \phi_{j+1}$
can be calculated in the initial state, then \eqref{e1} already gives
the exact prediction for the long-time limit of the 1-site EFP. This
could be used as a further check of the Diagonal Ensemble and the GGE, if the
time evolution could be simulated by independent numerical methods.

Also, the relation between the EFP and the charges deserves further
attention. It would be interesting to find a closed form result for
the $m$-site EFP with $m>2$.

\item In this work we demonstrated that the GGE is valid in the $q$-boson
  model. We argued that it holds for arbitrary $q$ and have rigorously
  proven it in the $q\to\infty$ limit in two special cases. 
The essential points were that the model has one particle
species, all charges have finite values, and the GETH could be proven directly.

One of  the most interesting open questions is whether there are other models
  where the GGE (built on the local charges only) gives a correct
  description of the stationary states.
\end{itemize}

We hope to return to these questions in further research.

\vspace{1cm}
{\bf Acknowledgements} 

\bigskip

We are grateful to G\'abor Tak\'acs and M\'arton Kormos for useful
discussions and for comments on the manuscript. In particular we are
thankful to  M\'arton Kormos for drawing our attention to the
$q$-boson model, and to G\'abor Tak\'acs for a fruitful discussion about
the symmetric polynomials relevant to the present work.

\bigskip

\addcontentsline{toc}{section}{References}
\bibliography{../../pozsi-general}
\bibliographystyle{utphys}

\end{document}